\documentclass[10pt,conference,compsocconf,letterpaper]{IEEEtran}

\IEEEoverridecommandlockouts % reenable \thanks command

\usepackage{amsmath,amsthm}
\usepackage{amssymb,latexsym}
\usepackage{amsfonts}
\usepackage{mathtools} % for text over arrows

\usepackage{hyperref} % for url

\usepackage{xcolor}
\newcommand{\Z}{\mathbb{Z}}							% Integer number
\usepackage{algorithm}
\usepackage{algorithmicx} % recommended by IEEEtran
\usepackage[noend]{algpseudocode} % pseudocode layout in algorithmicx

\algrenewcommand{\algorithmiccomment}[1]{\hfill {\color{gray} // #1}}
\algnewcommand{\lComment}[1]{\State {\color{gray} // #1}}

\makeatletter
\newcommand{\Setlineno}[1]{\setcounter{ALG@line}{\numexpr#1-1}}
\makeatother

\newcommand{\hStatex}[0]{\vspace{5pt}}

\newcommand{\lIf}[2]
{\State \algorithmicif\ {#1}\ \algorithmicthen\ {#2}}

\algnewcommand{\lElse}[1]
{\State \algorithmicelse\ {#1}}

\newcommand{\lIfElse}[3]
{\State \algorithmicif\ {#1}\ \algorithmicthen\ {#2} \algorithmicelse\ {#3}}

\usepackage{paralist} % for inparaenum

\newtheorem{theorem}{Theorem}[section]
\renewcommand{\IEEEQED}{\IEEEQEDopen}
\newtheorem{lemma}{Lemma}[section]
\newtheorem{definition}{Definition}[section]
\newtheorem{corollary}{Corollary}[section]
\usepackage{array} % for table environment
\newcommand{\innercell}[2]{\begin{tabular}{@{}#1@{}}#2\end{tabular}} 
\begin{document}

% paper title can use linebreaks \\ within to get better formatting as desired
\title{Verifying PRAM Consistency over Read/Write Traces of Data Replicas}

% author names and affiliations use a multiple column layout for up to three
% different affiliations

% \author{\IEEEauthorblockN{Hengfeng Wei$^{1,2}$, Yu Huang$^{1,2}$
% \IEEEauthorrefmark{1}}
%
% \thanks{\IEEEauthorrefmark{1}Corresponding author.}
%
% \IEEEauthorblockA{$^1$State Key Laboratory for Novel Software Technology \\
% Nanjing University, Nanjing 210093, China \\
% $^2$Institute of Computer Software, Nanjing University, Nanjing 210093, China \\
% hengxin0912@gmail.com, yuhuang@nju.edu.cn}}

\author{
\IEEEauthorblockN{Hengfeng Wei$^{1}$,
Marzio De Biasi$^{2}$, 
Yu Huang$^{1}$\IEEEauthorrefmark{1}, 
Jiannong Cao$^3$, Jian Lu$^{1}$}

\thanks{\IEEEauthorrefmark{1}Corresponding author.}

\IEEEauthorblockA{$^1$State Key Laboratory for Novel Software Technology\\
Nanjing University, Nanjing 210046, China\\
% $^2$Institute of Computer Software, Nanjing University, Nanjing 210046, China\\
hengxin0912@gmail.com, \{yuhuang, lj\}@nju.edu.cn\\
$^2$marziodebiasi@gmail.com\\
$^3$
% Internet and Mobile Computing Lab, Department of Computing\\
Hong Kong Polytechnic University, Hong Kong, China\\
csjcao@comp.polyu.edu.hk}
}

% make the title area
\maketitle

%--
\begin{abstract}
Data replication technologies enable efficient and highly-available data
access, thus gaining more and more interests in both the academia and the
industry. 
However, data replication introduces the problem of data consistency. 
Modern commercial data replication systems often provide weak consistency for
high availability under certain failure scenarios.
An important weak consistency is Pipelined-RAM (PRAM) consistency. 
It allows different processes to hold different views of data. 
To determine whether a data replication system indeed provides PRAM consistency,
we study the problem of \emph{Verifying PRAM Consistency over read/write traces}
(or VPC, for short).

We first identify four variants of VPC according to 
  \begin{inparaenum}[\itshape a\upshape)]
    \item whether there are Multiple shared variables (or one Single variable),
    \emph{and}
    \item whether write operations can assign Duplicate values (or only Unique
    values) for each shared variable;
  \end{inparaenum} the four variants are labeled VPC-SU, VPC-MU, VPC-SD, and
  VPC-MD.
Second, we present a simple VPC-MU algorithm, called \textsc{RW-Closure}. 
It constructs an operation graph $\mathcal{G}$ by iteratively adding edges
according to three rules.
% It models read/write operations as graph nodes, and precedence relations
% between operations as directed edges. 
% PRAM consistency is captured by three rules of iteratively adding edges to the
% transitive closure of the graph.
Its time complexity is $O(n^5)$, where $n$ is the number of operations in the
trace.
Third, we present an improved VPC-MU algorithm, called \textsc{Read-Centric},
with time complexity $O(n^4)$.
Basically it attempts to construct the operation graph $\mathcal{G}$ in an
\emph{incremental and efficient way}.
Its correctness is based on that of \mbox{\textsc{RW-Closure}}.
% Third, we present another VPC-MU algorithm, called \textsc{Read-Centric},
% with time complexity $O(n^4)$.
% It restricts the applications of the three rules in a Read-induced subgraph
% and manages them in a \emph{reverse topological order} of the subgraph.
Finally, we prove that \mbox{VPC-SD} (so is \mbox{VPC-MD}) is
\mbox{$\sf{NP}$-complete} by reducing the strongly \mbox{$\sf{NP}$-complete}
problem \mbox{\textsc{3-Partition}} to it.
\end{abstract}

%--

\IEEEpeerreviewmaketitle
% ============================
\section{Introduction}  \label{section:introduction}

Data replication consists of maintaining multiple copies of data,
called replicas, on separate computing entities. 
It is a critical enabling technology in distributed systems, improving
system performance, reliability, and scalability \cite{Vogels07,Cooper08,
Megastore11}.
Practically, it is desirable for a data replication system to achieve three
properties \emph{simultaneously}, namely data consistency (C), availability (A),
and partition-tolerance (P) \cite{Brewer00}.
However, this has been theoretically proved impossible by the CAP
theorem \cite{Gilbert02,Gilbert12}.
The impossibility result leads to multiple balance options, 
among which modern commercial data replication systems often
choose to sacrifice consistency under network partitions and certain failure
scenarios for high availability.
Thus, researchers have developed various weak consistency models such as
PRAM consistency (Pipelined RAM) \cite{Lipton88}, cache consistency
\cite{Goodman89} (a.k.a. memory coherence \cite{Li89}), causal consistency
\cite{Ahamad95}, processor consistency \cite{Steinke04}, and eventual consistency
\cite{Vogels09}, besides the strong ones such as linearizability
\cite{Herlihy90} (a.k.a. atomicity \cite{Lamport86}) and sequential consistency
\cite{Lamport79,Attiya94}.
For example, Yahoo!'s \mbox{PNUTS} \cite{Cooper08} provides
per-record timeline consistency (similar to the processor consistency). 
Amazon's Dynamo \cite{Vogels07} only promises eventual consistency. 
Nowadays, weak consistency is playing a more and more important role, with the
prevalence of cloud data storage services, mobile devices, and wireless
communications.

In this work, we focus on PRAM consistency \cite{Lipton88}, one
of the well-known weak consistency models.
Informally, a read/write trace satisfies PRAM consistency if and only if
write operations performed by a single process are observed by all the
other processes in the order they were issued, whereas write operations
from different processes may be observed in different orders by different
processes \cite{Steinke04}.
% PRAM consistency is weak in that it does not require all processes to agree on
% the same view of the order in which operations occur \cite{Attiya04}.
% PRAM consistency is weak in that write operations coming from different
% processes may be seen in a different order by different processes.
To illustrate its practical usefulness, let us consider the \emph{photo sharing
application} described in \cite{Cooper08}. In this application, users can post
photos and control their accesses. 
Now Alice wishes to share some photos with her classmates but not with her
mother. 
She does a sequence of updates to her album: adds her classmates to and
removes her mother from the album access list, and then posts photos. 
Under PRAM consistency, the updates from Alice are guaranteed to be seen by any
user in the order they were issued.

Different protocols can be designed to guarantee PRAM consistency. 
However, theoretically correct protocols can suffer from buggy implementations
and unexpected runtime failures.
Furthermore, the implementations of such systems, when they
are published as commercial web services, are often inaccessible to users.
Thus, the users can only test the system by observing and analyzing its logs
(i.e., read/write traces of operations) to verify whether it is delivering
promised consistency \cite{Golab11}.
% Therefore, it is desirable to verify whether the read/write traces of data
% replicas satisfy PRAM consistency as claimed by the service provider.
Though weak consistency models are regarded important, to the best of our
knowledge, their verification problems have not been sufficiently studied yet.
In this work, we systematically study the problem of \emph{verifying PRAM
consistency over read/write traces} (VPC, for short). Specifically,
\begin{itemize}
  \item First, we identify four variants of VPC according to 
  \begin{inparaenum}[\itshape a\upshape)]
    \item whether there are Multiple shared variables (or one Single variable),
    \emph{and}
    \item whether write operations can assign Duplicate values (or only Unique
    values) for each shared variable;
  \end{inparaenum} the four variants are labeled VPC-SU, VPC-MU, VPC-SD, and
  VPC-MD.
  \item Second, we present a simple VPC-MU algorithm, called \textsc{RW-Closure}.
  It constructs an operation graph $\mathcal{G}$ by iteratively adding edges
  according to three rules.
%   It models read/write operations as graph nodes, and precedence
%   relations between operations as directed edges. PRAM consistency is captured by
%   three rules of iteratively adding edges to the transitive closure of the
%   graph. 
  Its time complexity is $O(n^5)$, where $n$ is the number of
  operations in the trace.
  \item Third, we present an improved VPC-MU algorithm, called
  \textsc{Read-Centric}, with time complexity $O(n^4)$.
  Basically it attempts to construct the operation graph $\mathcal{G}$ in an
  \emph{incremental and efficient way}.
  It is incremental in that it processes, one at a time, the read operations.
  It is efficient because for each read operation, it applies the three rules 
  in a Read-induced subgraph and organize them in a reverse topological
  order of the subgraph.
  Its correctness is based on that of \textsc{RW-Closure}.
  \item Finally, we prove that VPC-SD (so is VPC-MD) is
  \mbox{$\sf{NP}$-complete} by reducing the strongly $\sf{NP}$-complete problem
  \textsc{3-Partition} \cite{Garey75, Garey79} to it.
\end{itemize}

The rest of this paper is organized as follows.
Section \ref{section:related_work} discusses the related work.
Section \ref{section:problem_definition} defines the problem of verifying PRAM
consistency over read/write traces and its four variants.
Sections \ref{section:rw_closure} and \ref{section:read_centric} present
the two \mbox{VPC-MU} algorithms: \textsc{RW-Closure} and
\textsc{Read-Centric} respectively.
Section \ref{section:npc} gives the $\sf{NP}$-completeness proof of VPC-SD (so
is VPC-MD).
Section \ref{section:conclusion} concludes the paper including suggestions
for future work.
% ============================
\section{Related Work} \label{section:related_work}

%%%%%%%%%%%%%%% related work on consistency conditions is deleted %%%%%%%%%%%%%%
% Many different levels of consistency conditions have been proposed, ranging
% from strong consistency such as linearizability (a.k.a. atomicity)
% \cite{Herlihy90, Lamport86} and sequential consistency \cite{Attiya94} to the
% weaker ones such as causal consistency \cite{Ahamad95}, PRAM consistency
% \cite{Lipton88}, and eventual consistency \cite{Vogels09}. 
% Relationships between consistency conditions are also explored.
% Specifically, Steinke and Nutt \cite{Steinke04} first identify a set of four
% consistency properties, and then reinterpret the existing consistency conditions
% as combinations of any subset of them. This way, they establishes a lattice
% structure of consistency conditions.
% Haldar and Vidyasankar \cite{Haldar07} identify five different types of
% ``recentness'' of the value a \emph{read} could return and use them to define
% five \emph{Read} illegalities. The consistency conditions are then
% reinterpreted in terms of the presence or absence of these \emph{Read}
% illegalities.
%   In addition, Attiya and Friedman define hybrid consistency to benefit both the
%   expressiveness of strong consistency conditions and the efficiency of weak
%   consistency conditions. Yu and Vahdat \cite{Yu02} introduce a continuous
%   consistency which allows for an tunable tradeoff between consistency and availability.

Many efforts have been made on the verification problems with respect to other
consistency models than PRAM.
In their seminal work, Gibbons and Korach \cite{Gibbons97} study the
verifying sequential consistency (VSC) and the verifying
linearizability (VL) problems. 
Both problems are proved to be $\sf{NP}$-complete in general. 
In addition, they define the VSC-read problem, in which a \emph{read-mapping} is
known, and prove that it remains \mbox{$\sf{NP}$-complete}. 
Here a read-mapping is a function mapping each read operation to a write
operation which was responsible for the value read.
% They also systematically study various variants of the two problems and
% provide a collection of complexity results.
Cantin et al. \cite{Cantin05} show that the verifying memory coherence
problem (VMC) is $\sf{NP}$-complete.
They also prove that the problem of verifying sequential consistency for
executions that are memory coherent (VSCC) remains \mbox{$\sf{NP}$-complete}. 
Golab et al. \cite{Golab11} study the verification problems with
respect to safety, atomicity, regularity, and sequential consistency. 
Beyond a yes/no answer, they seek online algorithms to detect a consistency
violation as soon as it appears. They also consider how to quantify the
severity of violations. 
More recently, Golab et al. \cite{Golab13} solve the verification problem of
$2$-atomicity (2-AV) and show that the weighted $k$-AV problem is
$\sf{NP}$-complete.
In this work we investigate the \emph{verifying PRAM consistency} (VPC) problem.
As far as we know, we are the first to systematically solve this problem.

In the context of shared memory multiprocessor, some relaxed memory consistency
models have been studied \cite{Hangal04, Roy06, Baswana08}.
%   The consistency models under consideration mainly focus on TSO (Total
%   Store Order) and RMO (Relaxed Memory Order) in which the program order
% between operations on the same processor is in some way allowed to be relaxed.
Specifically, Hangal et al. \cite{Hangal04} develop TSOtool to verify the
traces of programs against Total Store Order model when a read-mapping is known
(VTSO-read).
The time complexity of their algorithm is $O(n^5)$, where $n$ is the number of
operations in the trace.
% Manovit and Hangal \cite{Manovit05} further reduce the time complexity to
% $O(kn^3)$ for one of its variants (namely, VTSO-read), where $k$ is the number
% of processors.
Roy et al. \cite{Roy06} also deal with the VTSO-read problem and present a fully
parallelized algorithm with $O(n^4)$ time complexity.
Baswana et al. \cite{Baswana08} identify a graph problem called
implied-set-closure as the abstraction of the bottleneck of the VTSO-read
problem, and further reduce its time complexity to $O(n^3)$.
However, all the above algorithms only do approximate checking because
the problem itself is $\sf{NP}$-complete \cite{Hangal04}.
In contrast, we show that the VPC problem for traces in which write operations
do not assign duplicate values (thus a read-mapping is known) can be
\emph{completely} solved in polynomial time.
Although its basic idea is simple and resembles that of \cite{Hangal04}, its
correctness proof is one of our key contributions.
On the other hand, we prove its $\sf{NP}$-completeness for other traces.
% ============================
\section{Problem Definition} \label{section:problem_definition}

In this section, we first define read/write traces of data replicas and
PRAM consistency, and then define the problem of verifying PRAM consistency over
read/write traces.
%%%%%
\subsection{Read/Write Trace} \label{ss:rw_trace}

We model the data replicas as a collection of read/write shared variables
supporting read/write operations, and the separate computing entities as a
collection of processes.

\begin{definition} [Operation ($o$)] \label{def:operation}
  An operation $o$ is a quadruple $(t, p, v, d) \in \{R,W\} \times P \times V
  \times D$ where,
  \begin{itemize}
    \item $t \in \{R, W\}$ is the type of operation ($R$ for read and
    $W$ for write). 
    An operation is complete if a read has returned its value or a write has
    been acknowledged;
    \item $p \in P$ is the process issuing the operation;
    \item $v \in V$ is the variable to which the operation is applied;
    \item $d \in D$ is a valid value for the variable $v$.
  \end{itemize}
%   We assume that each process has at most one operation pending at a time.
%   Furthermore, each operation is completed instantaneously.
\end{definition}

We adopt the following notational conventions for operation $o = (t, p, v, d)$.
The process is denoted by $p(o)$. The variable and the value involved are
denoted by $var(o)$ and $val(o)$ respectively. 
Generally, we use $o$ for any operation, $r$ for any read operation, $w$ for any
write operation, $O$ for the set of all operations, $R$ for the set of all read
operations, $W$ for the set of all write operations, and $W_v$ for the set of
all write operations on the same variable $v$.

There are two basic partial orders between operations.
Program order, denoted $\prec_{PO}$, is the order in which operations are
issued by each process.
Write-to order, denoted $\prec_{WR}$, defines which write is read by each
read.

\begin{definition} [Program Order ($\prec_{PO}$)]
  $(o_1, o_2) \in \; \prec_{PO}$ if and only if $p(o_1) = p(o_2)$ and $o_1$ is
  issued (and completed) before $o_2$. 
  We employ $\preceq_{PO}$ to denote the reflexive closure of $\prec_{PO}$.
\end{definition}

\begin{definition} [Write-to Order ($\prec_{WR}$)]
  $(o_1, o_2) \in \; \prec_{WR}$ if and only if $o_1 \in W \land o_2 \in R$, and
  $var(o_1) = var(o_2) \land val(o_1) = val(o_2)$.
\end{definition}

We can now define the read/write traces as follows. 
\mbox{Figure~\ref{fig:closure_example}} in Section \ref{ss:closure_example}
shows an example of a read/write trace consisting of four processes.

\begin{definition} [Read/Write Trace ($T$)]
  A read/write trace $T$ of data replicas comprises multiple process
  histories, each of which consisting of a finite sequence of read and write
  operations in program order.
\end{definition}

%%%%%
\subsection{PRAM Consistency Model} \label{ss:pram_consistency}

The PRAM consistency model is one of the well-known weak consistency
models \cite{Lipton88, Steinke04}. 
It takes into account both program order and write-to order.
Informally, a \mbox{read/write} trace satisfies PRAM consistency if and
only if write operations performed by a single process are observed by all other
processes in the order they were issued (i.e., program order), whereas write
operations from different processes may be observed in different orders by
different processes \cite{Steinke04}.
There are two key points to explain.
First, PRAM consistency is weak in that it does not require
all the processes to agree on the same view of the order in which operations
occur. It implies that each process can be checked against PRAM
consistency separately.
Second, the operations \emph{visible} to each process $p$ are all write
operations and its own read operations, while ignoring read operations from
other processes (formally, it is the set of $\{o \mid (o \in W) \lor ( p(o)
= p \land o \in R )\}$).
Note that, for process $p$, its visible read operations are all on the same
process (i.e., $p$ itself). 

To state PRAM consistency formally, we first give some basic definitions on
\emph{schedule}.
A \emph{schedule} (denoted $\pi$) is just a sequence of operations. Given a
schedule, the \emph{precedence relation} between any two operations in it
is denoted by `$\prec$'. We employ $\preceq$ to denote the reflexive
closure of $\prec$. Moreover, we define $\min(o_1,o_2) = o_1$ and $\max(o_1,
o_2) = o_2$ if $o_1 \preceq o_2$.

A schedule $\pi$ of a set of operations $O$ is said to
\emph{respect} some partial order $\mathcal{P}$ (denoted $(\pi, \mathcal{P})$)
if and only if the schedule is a \emph{linearization} of the partial order. Formally,
  \[
    (\pi, \mathcal{P}) \iff \forall_{o_1, o_2 \in O} \big ((o_1, o_2) \in
    \mathcal{P} \Rightarrow o_1 \prec o_2 \big ).
  \]
Intuitively, the notion of \emph{respect} enforces a schedule to satisfy
specified partial orders.
Furthermore, the following notion of \emph{legal schedule} is considered a
fundamental correctness requirement for all consistency models \cite{Steinke04}.

\begin{definition} [Legal Schedule] \label{def:legal_schedule}
  A schedule $\pi$ of operations is \emph{legal} if and only if each read
  reads the value from the latest preceding write on the same variable
  in the schedule.
  Predicate $LS(\pi)$ is evaluated true if and only if the schedule $\pi$ is
  legal.
\end{definition}

% Furthermore, the notion of \emph{respect} enforces the schedule to satisfy
% specified partial orders.
% 
% \begin{definition} [Respect Property] \label{def:respect_property}
%   A schedule $\pi$ of a set of operations $O$ is said to \emph{respect} some
%   partial order $R$ (denoted $\pi^{R}$) if and only if the schedule is a
%   linearization of the partial order. Formally,
%   \[
%     \pi^{R} \iff \forall_{o_1, o_2 \in O} \big ((o_1, o_2) \in R \Rightarrow o_1
%     \prec o_2 \big ).
%   \]
% \end{definition}

\begin{definition} [PRAM Consistency] \label{def:pram_consistency}
  A read/write trace satisfies PRAM consistency if and only if for
  each process $p$, there exists a legal schedule $\pi$ of its visible
  operations, respecting both program order and write-to order. Formally,
  \[
    \forall_{p \in P}\; \exists_{\pi}\; \big(LS(\pi) \land
    (\pi, \prec_{PO} \cup \prec_{WR}) \big ).
  \]
\end{definition}

According to Definition \ref{def:pram_consistency}, we can verify each process
against PRAM consistency separately. In the remainder of this paper, we thus
focus on the verification problem with respect to some particular process and
distinguish it with $p_0$.
%%%%%
\subsection{The Problem of Verifying PRAM Consistency}
\label{subsection:problem}

The problem of \underline{V}erifying \underline{P}RAM \underline{C}onsistency
(VPC, for short) over \mbox{read/write} traces is defined as a decision problem.

\begin{definition}[Verifying PRAM Consistency Problem] \hfill
  \begin{itemize}
    \item \textbf{INSTANCE:} A read/write trace $T$. Its size (denoted $n$)
    is defined as the total number of the operations in it.
    \item \textbf{QUESTION:} Does $T$ satisfy PRAM consistency?
  \end{itemize}
\end{definition}

Following the terminology in \cite{Gibbons97}, we identify four variants
of the general VPC problem from two orthogonal dimensions:
\begin{inparaenum}[\itshape a\upshape)]
  \item whether there are Multiple shared variables (or one Single variable),
  \emph{and}
  \item whether write operations can assign Duplicate values (or only Unique
  values) for each shared variable.
\end{inparaenum}

%%%%%%%%%%%% table: variants of VPC %%%%%%%%%%
\begin{table}[!t]
  \renewcommand{\arraystretch}{1.5}%
   \caption{A summary of complexity results for VPC problem
   ($[\ast]: \textrm{new results}$).}
   \label{tbl:results}
  \centering
	\begin{tabular}{|c|c|c|}
      \hline
	  &	\it (S)ingle variable  & \it (M)ultiple variables  	
	  \\ \hline
      \it write (U)nique value &
      \innercell{c}{VPC-SU \\ (P) \cite{Golab11}} & 
      \innercell{c}{VPC-MU \\ (P) $[\ast]$} 
      \\ \hline 
      \it write (D)uplicate values &
      \innercell{c}{VPC-SD \\ (NPC) $[\ast]$} & 
      \innercell{c}{VPC-MD \\ (NPC) $[\ast]$} 
      \\ \hline
	\end{tabular}
\end{table}
%%%%%%%%%%%% table: variants of VPC %%%%%%%%%%

As summarized in Table \ref{tbl:results}, the VPC-SU variant can be solved
in polynomial time, following from \cite{Golab11}. 
In this paper, we address the other three variants. 
Specifically, we show that VPC-MU can also be solved in polynomial time by
presenting two algorithms: the \textsc{Read-Closure} algorithm with $O(n^5)$
time complexity and the \textsc{Read-Centric} algorithm with $O(n^4)$ time complexity. 
On the other hand, we prove that VPC-SD (so is VPC-MD) is $\sf{NP}$-complete by
reducing the strongly $\sf{NP}$-complete problem \textsc{3-Partition}
\cite{Garey75, Garey79} to it.

% ==========
\section{The \textsc{RW-Closure} Algorithm} \label{section:rw_closure}

In this section, we present a VPC-MU algorithm, called \textsc{RW-Closure}.
Note that in the trace of VPC-MU instance, for each read operation $r$, there
is at most one write (denoted $D(r)$ for dictating write) from which $r$ reads
the value. 
In practice, each write operation can be tagged with a globally unique
identifier, e.g., by combining its process id and a local sequence number
\cite{Golab11}.
% Thus, the assumption of writing unique values does not incur any loss of
% generality.
%%%%%
\subsection{Overview} \label{ss:closure_overview}

The \textsc{RW-Closure} algorithm models the read/write trace as a directed
graph with operations as nodes and precedence relations between operations as
directed edges.
PRAM consistency is captured by three kinds of edges.
The \textsc{RW-Closure} algorithm keeps adding such edges to the transitive closure
of the graph until no more edges can be added. Then the trace $T$ satisfies PRAM
consistency if and only if the resulting graph $\mathcal{G}$ is acyclic (i.e.,
DAG).

Specifically, at least two kinds of edges are necessary to meet PRAM
consistency: edges for program order and edges for write-to order.
The third kind of edges can be derived from the \emph{legal schedule} notion in
Definition \ref{def:legal_schedule} \cite{Steinke04, Roy06}.
In a legal schedule, between each read operation $r$ on variable $v$
and its dictating write operation $w = D(r)$, there cannot be any other
write (denoted $w'$) on the same variable $v$. This observation results
in two cases:
\begin{inparaenum}[\itshape 1\upshape)]
  \item if $w' \prec r$, we have $w' \prec w$; \emph{and}
  \item if $w \prec w'$, we have $r \prec w'$.
\end{inparaenum}
Thus we get the following four rules for adding edges in $\mathcal{G}$:
\begin{itemize}
  \item ({\bf Rule A: program order}) For any pair of operations $o_1$ and
  $o_2$, if $o_1 \prec_{PO} o_2$, then add an edge from $o_1$ to $o_2$.
  \item ({\bf Rule B: write-to order}) For any pair of operations $w$ and
  $r$, if $w \prec_{WR} r$, then add an edge from $w$ to $r$.
  \item ({\bf Rule C: w'wr order}) For any triple of operations $w, r$
  and $w'$ on the same variable, if $w = D(r) \land w' \prec r$, then
  add an edge from $w'$ to $w$, leading to \mbox{$w' \prec_{W'W} w \prec_{WR}
  r$}.
  Note that we denote the precedence relation between such $w'$ and $w$ by
  $\prec_{W'W}$.
  \item ({\bf Rule D: wrw' order}) For any triple of operations $w, r$ and
  $w'$ on the same variable, if $w = D(r) \land w \prec w'$, then add an edge
  from $r$ to $w'$, leading to $w \prec_{WR} r \prec_{RW'} w'$. 
\end{itemize}
As shown in the following, \emph{the first three rules} are sufficient for
the VPC-MU problem.
%%%%%%%%%%
\subsection{Detailed Design} \label{ss:closure_design}

In Algorithm~\ref{alg:closure}, Rule~A for program order edges and
Rule~B for write-to order edges are first applied (Lines
\ref{line:closure_ab_begin} - \ref{line:closure_ab_end}).
To apply Rule~C, it is expected to first identify the triples conformed to
it. To this end, the algorithm checks each pair of $r$ and $w = D(r)$,
and find out all potential $w'$ such that there is a path from $w'$ to $r$
(i.e., $w' \prec r$) (Lines~\ref{line:closure_rulec_begin} -
\ref{line:closure_rulec_end}).
The reachability relation between $w'$ and $r$ is computed by transitive closure
algorithm (Line~\ref{line:closure_ts}) based on an $n \times n$ Boolean
operation matrix (opMatrix).
If any edges are added by Rule~C, new triples conformed to Rule~C can
emerge due to updated reachability relation. Therefore, the algorithm keeps
applying Rule~C and computing the transitive closure, until no
more edges are added (Line~\ref{line:closure_iteration}).
Finally, it concludes that the trace satisfies PRAM consistency if the resulting
graph is acyclic (Line~\ref{line:closure_dag_true}).

%%%%%% alg: \textsc{RW-Closure} algorithm %%%%%
% \begin{figure}[!t]
\begin{algorithm}[!t]
  \caption{The \textsc{RW-Closure} algorithm.}
  \label{alg:closure}
  \begin{algorithmic}[1]
%   \Require: a read/write trace $T$.
%   \Ensure: true, if $T$ satisfies PRAM consistency; false, otherwise.
% 	\lComment{a read/write trace $T$.}
% 	\lComment{true, if $T$ satisfies PRAM consistency; false, otherwise.}
    \State {\bf apply} Rule A to add edges for program order 
    \label{line:closure_po} \label{line:closure_ab_begin}
    \State {\bf apply} Rule B to add edges for write-to order
	\lIf{$\exists r (r \textrm{ {\it has no} } D(r))$}{\Return false}
    \label{line:closure_w2} \label{line:closure_ab_end}
    
    \hStatex
    \State {{\bf compute} the transitive closure of $\mathcal{G}$}
    \label{line:closure_ts} \label{line:closure_step3_begin}
%     \lComment{apply Rule C to add edges for $w'wr$ order:}
    
    \ForAll{\emph{read operation} $r$ \emph{in program order}}
      \label{line:closure_rulec_begin}
      \State $w \gets D(r), v \gets var(r)$
      \ForAll{$w' \neq w$ s.t., $\textrm{opMatrix}[w'][r] = 1$}
        \If{$var(w') = v \land \textrm{ opMatrix}[w'][w] = 0$}
		  \State $\textrm{opMatrix}[w'][w] \leftarrow 1$ \label{line:step3_end}
		\EndIf
      \EndFor
    \EndFor
    \label{line:closure_rulec_end} \label{line:closure_step3_end}
    
    \lIf{\emph{any edges are added by Rule~C}}{goto Line
    \ref{line:closure_step3_begin}}
    \label{line:closure_iteration}
    
    \hStatex
    \lIfElse{$\mathcal{G}$ \emph{is a DAG}}{\Return true}{\Return false}
    \label{line:closure_dag_true}
    
  \end{algorithmic}
\end{algorithm}
% \end{figure}
%%%%%% end alg %%%%%%%%
\subsection{An Illustrating Example} \label{ss:closure_example}

\begin{figure}[t]
  \centering
	  \includegraphics[width=0.48\textwidth]{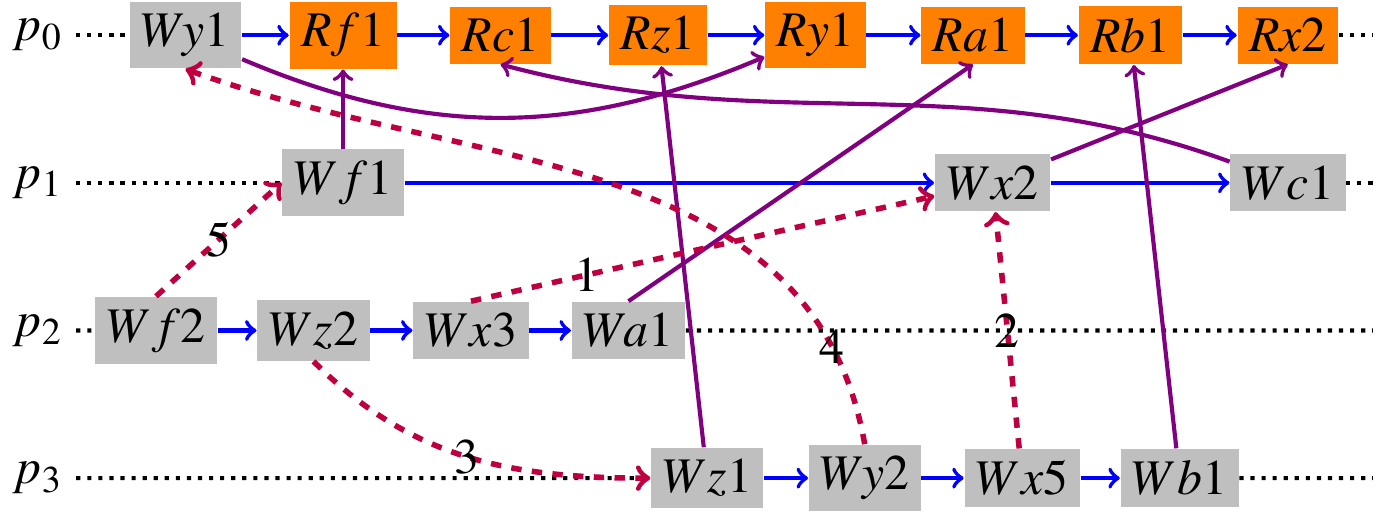}
	  \caption{Repeatedly applying Rule C to the transitive closure of
	  the operation graph in Algorithm~\ref{alg:closure}.}
	  \label{fig:closure_example}
  \vspace*{-10pt}
\end{figure}

Figure~\ref{fig:closure_example} shows a running example for the \textsc{RW-Closure}
algorithm.
The edges for program order and write-to order are denoted by solid lines. The
edges added by Rule~C are denoted by dashed lines, with labels indicating the
order in which they are added. 
Note that after the application of Rule~C to triple $Wy2, Wy1, \textrm{and }
Ry1$ (label 4), a new path from $Wf2$ to $Rf1$ arises (via edges with label 3
and label 4), and leads to another application of Rule~C to triple $Wf2, Wf1,
\textrm{and } Rf1$ (label 5).

We can figure out a \emph{legal schedule} of all the operations as a witness to
PRAM consistency (Equation. \ref{eq:schedule}). Note that the read operations are
bold and are separated by semicolons.
  \begin{align} \label{eq:schedule}
    Wf2\; Wf1\; Wz2\; Wz1\; Wy2\; Wy1\; \textbf{Rf1}; \nonumber \\
    Wx5\; Wx3\; Wx2\; Wc1\; \textbf{Rc1}; \textbf{Rz1}; \textbf{Ry1};\\
    Wa1\; \textbf{Ra1}; Wb1\; \textbf{Rb1}; \textbf{Rx2} \nonumber.
  \end{align}
%%%%%
\subsection{Correctness Proof} \label{ss:closure_correctness}

If the resulting graph $\mathcal{G}$ of Algorithm~\ref{alg:closure} is a DAG,
we expect to construct some \emph{legal schedule} (denoted $\pi_{\mathcal{G}}$)
as a witness to PRAM consistency.
To this end, a \emph{specific topological sorting} on $\mathcal{G}$ is
performed. It is based on the following two notations.

Intuitively, $r\textrm{-}$downset consists of all the operations which must
be scheduled before $r$, plus $r$ itself.

\begin{definition}[$r\textrm{-}$downset ($r_\Downarrow$)] \label{def:r_downset}
	$r\textrm{-}$downset of a read operation $r$ is a set $r_{\Downarrow}$
	of operations such that,
	\begin{itemize}
	  \item $r \in r_{\Downarrow}$;
	  \item $o \in r_{\Downarrow} \land o' \prec o \Rightarrow o' \in
	  r_{\Downarrow}$.
	\end{itemize}
\end{definition}

Let $r$ be a read operation and $r'$ be $r$'s previous read operation.
We use $r\textrm{-}$delta to refer to the ``extra'' operations
which are also scheduled before $r$, besides those in $r'\textrm{-}$downset.
In other words, $r\textrm{-}$delta (denoted $r_{\delta}$) of a read operation $r$
is a set (of operations) which equals the relative complement of
$r'_{\Downarrow}$ with respect to $r_{\Downarrow}$ (i.e., $r_{\Downarrow} \setminus r'_{\Downarrow}$).
For the first read operation $r$ on process $p_0$, we define $r_{\delta} =
r_{\Downarrow}$.
% \begin{definition} [$r\textrm{-}$delta ($r_{\delta}$)] \label{def:r_delta}
%   $r\textrm{-}$delta of a read operation $r$ is a set $r_{\delta}$ of
%   operations which equals the relative complement of $r'_{\Downarrow}$ with
%   respect to $r_{\Downarrow}$ (i.e., $r_{\Downarrow} \setminus
% r'_{\Downarrow}$).
%   For the first read operation $r$ on process $p_0$, we define $r_{\delta} =
%   r_{\Downarrow}$.
% \end{definition}
In terms of $r\textrm{-}$delta, we can now describe the construction of the
legal schedule $\pi_{\mathcal{G}}$.

\begin{definition} [DAG-schedule ($\pi_{\mathcal{G}}$)] \label{def:dag_schedule}
  Given the resulting DAG $\mathcal{G}$ of Algorithm~\ref{alg:closure}, the
  legal schedule $\pi_{\mathcal{G}}$ (initially, it is an empty sequence) is
  constructed as follows:
  \begin{itemize}
%     \item Initialize $\pi_{\mathcal{G}}$ to be the empty sequence;
    \item Repeatedly take each read operation $r$ on process $p_0$ in program
    order, perform any topological sorting on $r_{\delta}\textrm{-}$induced
    subgraph, and append it to $\pi_{\mathcal{G}}$.
  \end{itemize}
\end{definition}

The example in Section \ref{ss:closure_example} gives an illustration of such
schedule (Equation \ref{eq:schedule}). 
% Note that read operations are separated by semicolons.

\begin{lemma} \label{lemma:legal_schedule}
  If the resulting graph $\mathcal{G}$ of Algorithm~\ref{alg:closure} is
  acyclic, the schedule $\pi_{\mathcal{G}}$ constructed in
  Definition~\ref{def:dag_schedule} is legal.
\end{lemma}

\begin{figure}[!t]
  \centering
  \includegraphics[width = 0.48\textwidth]{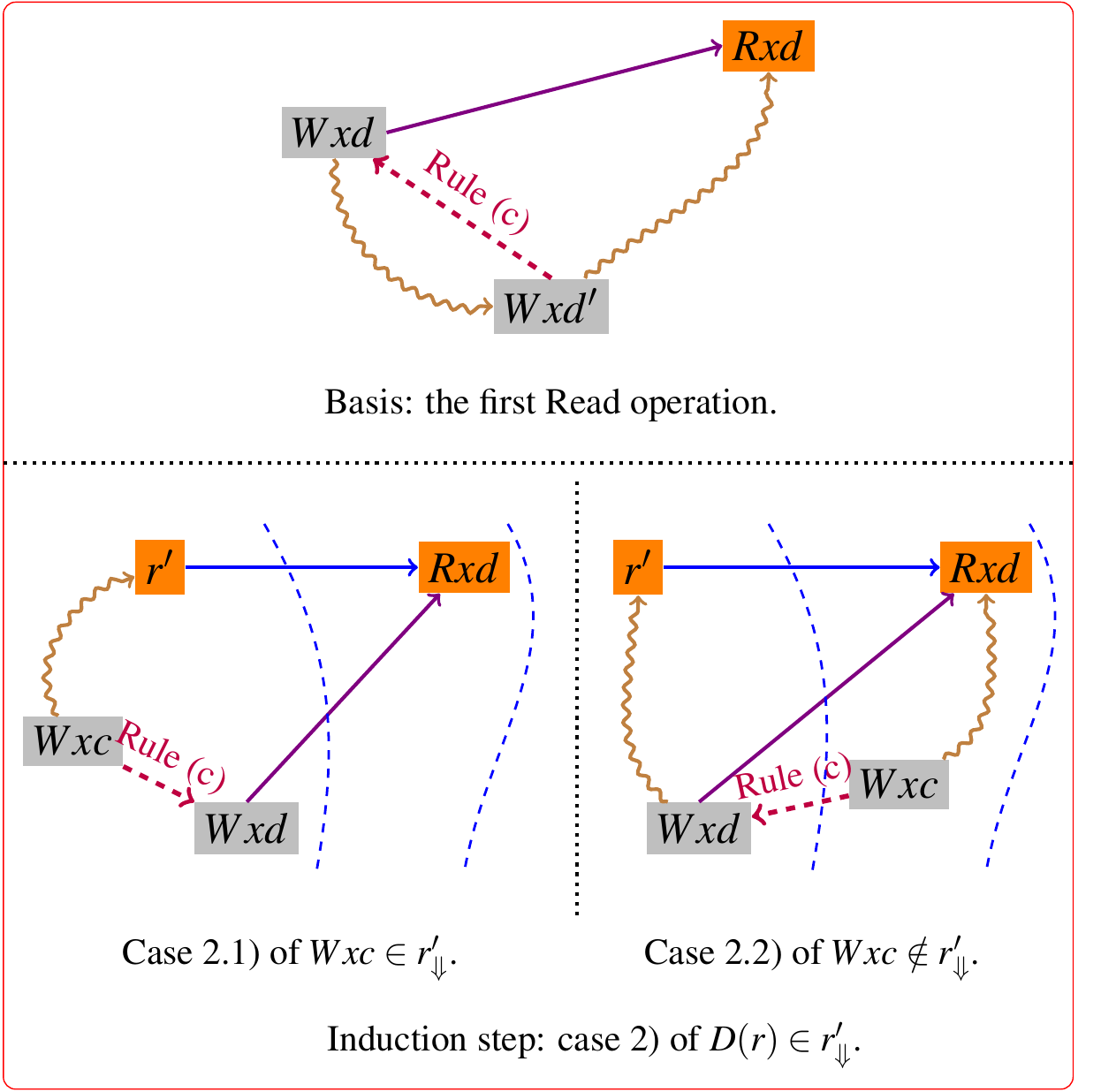}
  \caption{Correctness proof of Lemma \ref{lemma:legal_schedule}: $Rxd$ can be
  legally scheduled according to Definition \ref{def:dag_schedule}.}
  \label{fig:dag_schedule}
\end{figure}

\begin{IEEEproof}
  We prove this lemma by induction on the read operations on process $p_0$ in
  program order.
  
\emph{(Basis)} For the first read operation $r = Rxd$ and its
$r\textrm{-}$downset $r_{\Downarrow}$,
  
  \begin{itemize}
    \item It could not be the case that $Rxd \prec_{PO} Wxd$;
	\item Its dictating write operation $D(r) = Wxd$ could not be overwritten, say,
	by $Wxd'$. Otherwise $Wxd$ and $Wxd'$ create a cycle (Figure~\ref{fig:dag_schedule}); 
  \end{itemize}
  
Thus, any topological sorting on the $r_{\Downarrow}\textrm{-}$induced subgraph
is a (sub) legal schedule for $Rxd$.
		
\emph{(Induction hypothesis)} Assuming that the first $(n-1)$ read operations
have been legally scheduled according to Definition \ref{def:dag_schedule}, it
remains to prove that the $n^{th}$ read operation (denoted $r = Rxd$) will be
legally scheduled in the same way. Let $r'$ be the $(n-1)^{th}$ read operation.
		
\emph{(Induction step)} There are two cases according to whether $r$'s dictating
write operation $D(r) = Wxd$ has been scheduled before (i.e., $D(r) \in
r'_{\Downarrow}$).
		
\emph{1) $(D(r) \notin r'_{\Downarrow})$} By a similar argument to that of
\emph{Basis}, $Wxd$ would not be overwritten. And any topological sorting on the
$r_{\delta}\textrm{-}$induced subgraph does not break its legality of
the existing (sub) schedule.
Thus, we can append it to the existing schedule to obtain a legal one for
the first $n$ read operations.

\emph{2) $(D(r) \in r'_{\Downarrow})$} We show that $Wxd$ would not be
overwritten by write operations in $r_{\Downarrow}$, say, $Wxc$ (Figure
\ref{fig:dag_schedule}).

\emph{2.1) If $Wxc \in r'_{\Downarrow}$}, then we have 
\[
  Wxc \in r'_{\Downarrow} \land r' \prec_{PO} Rxd \Rightarrow Wxc \prec Rxd,
  \textrm{ and}
\]
\[
  Wxc \prec Rxd \land Wxd \prec_{WR} Rxd \xRightarrow{\text{Rule C}} Wxc \prec
  Wxd.
\]

\emph{2.2) If $Wxc \notin r'_{\Downarrow}$}, we show that $Wxc \notin
r_{\delta}$ either by contradiction:
\[
  Wxc \prec Rxd \land Wxd \prec_{WR} Rxd \xRightarrow{\text{Rule C}} Wxc
  \prec Wxd, \textrm{ and}
\]
\[
  Wxc \prec Wxd \land Wxd \in r'_{\Downarrow} \xRightarrow{\text{Definition
  \ref{def:r_downset}}} Wxc \in r'_{\Downarrow}.
\]
Thus, by performing any topological sorting on the $r_{\delta}\textrm{-}$
induced subgraph, and appending it to the existing schedule, we obtain a
legal one for the first $n$ read operations. 		 	
\end{IEEEproof}

The correctness of the \textsc{RW-Closure} algorithm is stated in the
following theorem.

\begin{theorem} \label{thm:closure_correctness}
  The VPC-MU instance satisfies PRAM consistency if and only if the resulting
  graph $\mathcal{G}$ of the \mbox{\textsc{RW-Closure}} algorithm is acyclic.
\end{theorem}

\begin{IEEEproof}
  ($\Rightarrow$) \emph{By contradiction}. If the resulting graph
  $\mathcal{G}$ is not a DAG, there exists some operation scheduled before
  itself.
	
  ($\Leftarrow$) \emph{If the resulting graph $\mathcal{G}$ is acyclic},
  Lemma \ref{lemma:legal_schedule} shows that the schedule $\pi_{\mathcal{G}}$
  constructed in Definition \ref{def:dag_schedule} is legal.
\end{IEEEproof}

% As mentioned in Section \ref{ss:closure_overview}, Rule (d) (i.e., $wrw'$
% order) is not necessary for the VPC-MU problem. We now explain this point more
% directly.
% In Figure~\ref{fig:closure_proof} (3), $R, W, W'$ are all on the same variable.
% $R$ reads the value from $W$ and $W$ precedes $W'$. It is expected to show that
% the application of Rule (d) to $W, R, \textrm{and } W'$ is dispensable in the sense
% of cycle detection.
% In other words, if there is a cycle involving the edge $R \to W'$, there will
% also be another cycle without it.
% First, it is necessary for $W' \prec R$ (edge with label 2) to close the
% cycle involving $R \to W'$. Consequently, $W' \prec_{W'W} W$ (edge with
% label 3), due to Rule~C, closes another cycle together with $W \prec W'$ (edge
% with label 1).
%%%%%
\subsection{Time and Space Complexity}  \label{ss:closure_complexity}

The worst-case time complexity of the \textsc{RW-Closure} algorithm is
dominated by the cost for Step 3 (Lines~\ref{line:closure_step3_begin} -
\ref{line:closure_step3_end} in Algorithm~\ref{alg:closure}).
The transitive closure of $\mathcal{G}$ (Line
\ref{line:closure_ts}) can be computed in $\Theta(n^3)$ time using
Floyd-Warshall's algorithm \cite{Cormen09}.
Applying Rule~C costs $O(n^2)$ to explores potential pairs of
nodes (Lines~\ref{line:closure_rulec_begin} - \ref{line:closure_rulec_end}).
The iteration over Step 3 and Step 4 may loop at most $O(n^2)$ times,
adding one edge by Rule~C in each iteration. 
In total, the worst-case time complexity of the \textsc{RW-Closure} algorithm
is $O(n^5)$.

Its space complexity is $\Theta(n^2)$, for the Boolean operation matrix
(opMatrix).
%%%%%%%%%%%%%%%%%%%%%%%%%%%%%%%%%%%%%%%%%%%%%%%%%%%%%%%%%%%%%%%%%%%%%%%%%%%%%%%%
\section{The \textsc{Read-Centric} Algorithm} \label{section:read_centric}

In this section, we present an improved VPC-SD algorithm, called
\textsc{Read-Centric}, with worst-case time complexity $O(n^4)$.
Its correctness proof is based on the previous \mbox{\textsc{RW-Closure}}
algorithm.
% It has two advantages over the \textsc{RW-Closure} algorithm. 
% First, its worst-case time complexity is $O(n^4)$. 
% Second, it is able to locate the first read operation which violates PRAM
% consistency.
%%%%%%%%%%%%%%%
\subsection{Overview} \label{ss:centric_overview}

In Theorem \ref{thm:closure_correctness}, we have shown that the trace $T$
satisfies PRAM consistency if and only if the resulting graph $\mathcal{G}$ of the
\textsc{RW-Closure} algorithm is acyclic. 
Generally speaking, the \mbox{\textsc{Read-Centric}} algorithm attempts to
construct graph $\mathcal{G}$ in an incremental and efficient way.
It is incremental in that it processes the read operations
on process $p_0$ \emph{sequentially}.
It is efficient because for each read operation, it applies Rule~C
\emph{locally and in a well-organized order}. 
Algorithm~\ref{alg:read_centric} sketches its basic idea. 

Let $r$ be the current read operation under scrutiny, $r'$ be $r$'s
previous read operation, and $v$ be the variable of $r$.
Upon read operation $r$, the \textsc{Read-Centric} algorithm
first initializes the reachability relation concerning the incrementally new
operations in $r_{\delta} = r_{\Downarrow} \setminus r'_{\Downarrow}$
% first identifies its $r\textrm{-}$downset $r_{\Downarrow}$, which comprises all
% the operations which must be scheduled before $r$ \emph{known till that time}
(Line \ref{line:centric_init_reachability}). 
(Here both $r_{\Downarrow}$ and $r'_{\Downarrow}$ are obtained according to
Definition \ref{def:r_downset} with respect to the dynamic graph $\mathcal{G}$
\emph{till that time}.)
It then attempts to \emph{schedule} \emph{locally} on the
$r_{\Downarrow}\textrm{-}$induced subgraph.
Specifically, the \emph{schedule} procedure starts with a simple observation
that $r$ must read from its dictating write operation $D(r)$ (Lines
\ref{line:centric_start_schedule_begin} - \ref{line:centric_start_schedule_end}).
According to Rule~C, any write operation $w'$ in $r\textrm{-}$downset on the
variable $v$ other than $D(r)$ must be scheduled before $D(r)$. 
% Here a write operation is \emph{active} means that it has not been overwritten. 
Thus the edges like $w' \to D(r)$ are added, updating the reachability relation
between operations.
Consequently, more applications of Rule~C may be triggered.
There are two cases to consider:
\begin{inparaenum}[\itshape 1\upshape)]
  \item $D(r) \notin r'_{\Downarrow}$ 
  \emph{and}
  \item $D(r) \in r'_{\Downarrow}$.
\end{inparaenum}
In the former case (Line~\ref{line:centric_case1}), the new added edges like $w'
\to D(r)$ have no effect on the reachability relation between the operations
from $r'_{\Downarrow}$. 
% The \emph{schedule} procedure simply returns. 
In the latter one (Line~\ref{line:centric_call_schedule}), the operations in
$r_{\Downarrow}$ should be locally scheduled. 
This involves a serial of applications of Rule~C.
Contrast to that of the \textsc{RW-Closure} algorithm, the applications of
Rule~C here are carried out in a \emph{reverse topological order} of
the $r_{\Downarrow}$-induced subgraph.
Once some cycle is created, the algorithm aborts and outputs ``no''. 
If all the read operations are processed and no cycles arise, the algorithm
terminates and outputs ``yes''.
% Note that we have ruled out two trivial violations of PRAM consistency (Lines
% \ref{line:centric_preprocessing_begin} - \ref{line:centric_preprocessing_end}):
% \begin{inparaenum}[\itshape 1\upshape)]
%   \item some read operation $r$ has no dictating write operation $D(r)$;
%   \emph{or}
%   \item $r$ happens before $D(r)$ in program order.
% \end{inparaenum}

%%%%%%%%% alg: read-centric (basic idea) %%%%%%%%%
% \begin{figure}[!t]
\begin{algorithm}[!t]
  \caption{The \textsc{Read-Centric} algorithm (sketch).}
  \label{alg:read_centric}
  \begin{algorithmic}[1]
    \State {\bf apply} Rule A to add edges for program order
    \label{line:centric_preprocessing_begin}
	\State {\bf apply} Rule B to add edges for write-to order
% 	\lComment{rule out two trivial cases}
	\lIf{$\exists r (r \textrm{ {\it has no} } D(r) \lor r \prec_{PO}
	D(r))$}{\Return false}
	\label{line:centric_preprocessing_end}
	
	\hStatex
	\ForAll{\emph{read operation} $r$ \emph{in program order}}
	\label{line:centric_foreach}
	\label{line:centric_body_begin}
	  \State Let $r'$ be $r$'s previous read operation
	  \State $v \gets var(r)$
	  
 	  \hStatex
% 	  \lComment{initialize reachability concerning
% 	  $r_{\delta} = r_{\Downarrow} \setminus r'_{\Downarrow}$}
	  \State \Call {Init-Reachability}{$r',r$}
	  \label{line:centric_init_reachability}
	  
 	  \hStatex
% 	  \lComment{start the schedule with $r$ reading from $D(r)$}
	  \ForAll{$w' \textrm{ {\it s.t.,} } w' \neq D(r) \land w' \in \textsl{LW}[v]$}
	    \label{line:centric_start_schedule_begin}
        \State add edge $w' \to D(r)$
        \If{\Call{Cycle-Detection}{$w',D(r)$}}
          \State \Return false
        \EndIf
        \State \Call{Update-Reachability}{$w',D(r),r$}
	  \EndFor
	  \label{line:centric_start_schedule_end}
	  
%  	  \hStatex
	  \vspace{4pt}
% 	  \lComment{case 1): reachability relation does not change}
	  \lIf{$D(r) \notin r'_{\Downarrow}$}{{\bf continue}} \Comment{case 1)}
	  \label{line:centric_case1}
	  
% 	  \lComment{case 2): locally schedule} 
	  \State $cycle \gets$ \Call{Topo-Schedule}{$r_{\Downarrow}$} \Comment{case 2)}
 	  \label{line:centric_call_schedule} 
	  \lIf{$cycle$}{\Return false}
	\EndFor
	\State \Return true
	\label{line:centric_body_end}
	   
  \end{algorithmic}
\end{algorithm}
% \end{figure}
%%%%%%%% end alg %%%%%%%%

Before describing the algorithm in detail, we first introduce some terminology
and notations.

%%%%%%%%%%%
\subsection{Terminology and Notations} \label{ss:centric_terminology}

During the course of \textsc{Topo-Schedule}, the
$r_{\Downarrow}\textrm{-}$induced subgraph is dynamic in that edges are added on
demand due to Rule~C.
To capture the dynamic reachability relation, two kinds of information are
dynamically maintained.

% Initially \textsl{LW} associates each variable with an empty set. It
% supports the standard dictionary procedures \textsc{LW-Lookup($v$)}
% ($\textsl{LW}[v]$, for short) and \textsc{LW-Reassign($v, LW_v$)}
% ($\textsl{LW}[v] \gets LW_v$, for short).

% For convenience, we provide another procedure \textsc{AW-Deactivate} to
% deactivate specified write operations for specified variable.

% %%%%%%% procedure Deactive of AW
% \begin{algorithmic}[0]
%   \Procedure{AW-Deactive}{$v, W_v$} 
%   \lComment{the write operations in $W_v$ are on the variable $v$}
%   \State $old\_W_{v} \gets \textsl{AW}[v]$
%   \State $new\_W_{v} \gets old\_W_{v} \setminus W_v$
%   \State $\textsl{AW}[v] \gets new\_W_{v}$
%   \EndProcedure
% \end{algorithmic}
% %%%%%%%% 

First, \textsl{ReachableRead} maintains, for each write operation, the
\emph{first} read operation it can reach via the precedence relation (i.e., $\prec$). 
Recall that read operations are all program ordered on the process $p_0$.
% We assign each read operation $r$ its ordinal $ord(r) \in \{1,2,\ldots,|R|\}$ in
% the total order. read operation with ordinal $i$ is denoted $R[i]$.

\begin{definition}[\textsl{ReachableRead (RR)}] \label{def:reachableread}
  \textsl{ReachableRead} is a \emph{dictionary} composed of a collection of $(w,
  r) \in W \times R$ pairs such that 
  \[
    \textsl{RR}[w] = r \Leftrightarrow w \prec r \land \nexists_{r' \prec_{PO}
    r} (w \prec r').
  \]
\end{definition}

% Initially, \textsl{ReachableRead} associates each write operation with a
% dummy read operation (e.g., \textsl{NILREAD}). It supports the standard
% dictionary procedures \textsc{RR-Lookup($w$)} (\textsl{RR}$[w]$, for short) and
% \textsc{RR-Reassign($w,r$)} (\textsl{RR}$[w] \gets r$).

Being complementary to \textsl{ReachableRead}, \textsl{PrecedingWrite}
maintains, for each operation, the \emph{last} write operation for each
variable preceding it. Strictly speaking,

\begin{definition}[\textsl{PrecedingWrite} (PW)] \label{def:precedingwrite}
  \textsl{PrecedingWrite} is a \emph{two-dimensional dictionary}.
  For each operation $o$, $\textsl{PW}[o]$ is a \emph{dictionary} composed of a
  collection of $(v,w) \in V \times W$ pairs with the following conditions:
  \begin{enumerate}
    \item $w \prec o \land var(w) = v$;
    \item $\exists_{r \in R} \; (w \prec_{WR} r)$;
    \item for any other $w'$ satisfying 1) - 2), we have $w' \prec w$.
  \end{enumerate}
\end{definition}

Condition 1) focuses on the preceding write operations on the same variable $v$.
Condition 2) concerns only the ones which have dictated read operations.
Condition 3) requires that all write operations satisfying 1) - 2) be totally
ordered. This is justified due to Rule~C and the fact that all read operations
are program ordered. Moreover, the precedence relation between them is
determined by the program order of their respectively \emph{first} dictated read
operations.

% \textsl{PrecedingWrite} supports the standard dictionary procedures
% \textsc{PW-Lookup($o,v$)} ($\textsl{PW}[o][v]$, for short) and 
% \mbox{\textsc{PW-Reassign($o,v,w$)}} ($\textsl{PW}[o][v] \gets w$, for short). 
Initially, \textsl{PrecedingWrite} associates each $\textsl{PW}[o][v]$ with a
dummy write operation \textsl{NILWRITE} which precedes all write operations.
It provides procedure \mbox{\textsc{PW-Update($o,o'$)}} to
update $\textsl{PW}[o']$ based on $\textsl{PW}[o]$ for each variable.

%%%%%%%%%%%%%%%%%%% procedure: PW-Update
\begin{algorithm}[!h]
\begin{algorithmic}[1]
  \Procedure{PW-Update}{$o,o'$}
  \ForAll{$v \in V$} \Comment{update to the latter write}
    \State $\textsl{PW}[o'][v] \gets \max(\textsl{PW}[o'][v],
    \textsl{PW}[o][v])$
  \EndFor
%   \lComment{consider $o$ now}
  \If{$o$ {\it has dictated read operations}}  \Comment{consider $o$}
	\State $\textsl{PW}[o'][var(o)] \gets o$
  \EndIf
  \EndProcedure
\end{algorithmic}
\end{algorithm}
% \end{figure}
%%%%%%%%%%%%%%%%%%%

Both \textsl{ReachableRead} and \textsl{PrecedingWrite} are used in
procedure \textsc{Apply-Rule-C} 
(more specifically, in its sub-procedures \textsc{Identify-Rule-C} and
\textsc{Cycle-Detection}, respectively).
They are updated once some edge is added.

Besides, we use \textsl{LocalWrites} to simply keep record of, for each
variable, the write operations locally in $r_{\Downarrow}$. Formally, 
\textsl{LocalWrites} (denoted \textsl{LW}) is a dictionary composed of a
collection of $(v, LW_{v}) \in V \times 2^{W_v}$ pairs. 
Recall that $W_v$ stands for the set of write operations on the same variable
$v$.

% \begin{definition}[\textsl{LocalWrites (LW)}] \label{def:localwrites}
%   \textsl{LocalWrites} is a \emph{dictionary} composed of a collection of
%   $(v, LW_{v}) \in V \times 2^{W_v}$ pairs. Recall that $W_v$ stands for
%   the set of write operations on the same variable $v$.
% \end{definition}

%%%%%
\subsection{Detailed Design} \label{ss:centric_design}

In this section, we first describe \mbox{\textsc{Init-Reachability}}
(called in Line~\ref{line:centric_init_reachability} of
Algorithm~\ref{alg:read_centric}) preparing for the key procedure
\textsc{Topo-Schedule}.
We then describe procedure \mbox{\textsc{Apply-Rule-C}} and its sub-procedures.
% namely \mbox{\textsc{Identify-Rule-C}}, \mbox{\textsc{Cycle-Detection}}, and
% \mbox{\textsc{Update-Reachability}}.
Particularly, during the course of \textsc{Topo-Schedule} we will
show how to perform \textsc{Apply-Rule-C} locally and in a well-organized order.

\subsubsection{Procedure \textsc{Init-Reachability}} Upon each read operation
$r$ and its previous read operation $r'$, the procedure
\textsc{Init-Reachability} initializes the reachability relation,
in terms of \mbox{\textsl{ReachableRead}} and \mbox{\textsl{PrecedingWrite}},
concerning the operations in $r_{\delta} = r_{\Downarrow} \setminus
r'_{\Downarrow}$ (Algorithm~\ref{alg:init_reachability}).
Here both $r_{\Downarrow}$ and $r'_{\Downarrow}$ are obtained according to
Definition \ref{def:r_downset} with respect to the dynamic graph $\mathcal{G}$
\emph{till the time} when \mbox{\textsc{Init-Reachability}} is called.
On the one hand, the first reachable read operation (i.e., \mbox{\textsl{RR}})
of each write operation in $r_{\delta}$ is now $r$ (Line
\ref{line:setup_reachability_rr}).
On the other hand, we initialize $\textsl{PW}$ of each operation in
program order.
Specifically, the operations in $r_{\delta}$ (except $r$) are partitioned
into two groups (both could be empty):
\begin{inparaenum}[\itshape 1\upshape)]
  \item the $rr\textrm{-}$group (denoted $grp_{rr}$) consists of all the write
  operations between $r'$ and $r$ on process $p_0$ (both exclusive); \emph{and}
  \item the $ww\textrm{-}$group (denoted $grp_{ww}$) consists of the rest on
  the same process with that of $D(r)$.
\end{inparaenum} 
Both groups are scanned through to initialize the \textsl{PW} of each operation
in the same manner (Lines~\ref{line:init_rr_group_begin} -
\ref{line:init_ww_group_end}).
% Second, the operations in $ww\textrm{-}group$ must be scheduled after the last
% write operation $lw$ which is in $r'_{\Downarrow}$ and is on the same process
% with that of $D(r)$ (Lines~\ref{line:aw_ww_group_begin} -
% \ref{line:aw_ww_group_end}).

%%%%%%%%%%%% procedure: Compute-ActiveWrites
\begin{algorithm}[!t]
  \caption{Procedure \textsc{Init-Reachability}.}
  \label{alg:init_reachability}  
  \begin{algorithmic}[1]
	\Procedure{Init-Reachability}{$r',r$}

    \ForAll{\emph{write operation} $w \in r_{\delta} = r_{\Downarrow}
    \setminus r'_{\Downarrow}$}
 	  \State $\textsl{RR}[w] \gets r$ \label{line:setup_reachability_rr}
%  	  \Comment{$w$'s first reachable read is now $r$} 
	  \State $v \gets var(w)$; $\textsl{LW}[v] \gets \textsl{LW}[v] \cup \{ w \}$ 
	  \Comment{collect $w$}
    \EndFor
    
    \hStatex
% 	\lComment{1) scan through $rr\textrm{-}group:$}
	\State $o_{pre} \gets r'$
	\label{line:init_rr_group_begin}
	\ForAll{$w \in grp_{rr}$}
 	  \State \Call{PW-Update}{$o_{pre}, w$}; $o_{pre} \gets w$ 
%  	  \State $o_{pre} \gets w$
	\EndFor
 	\State \Call{PW-Update}{$o_{pre}, r$} 
 	\Comment{update \textsl{PW} of $r$}
  	\label{line:init_rr_group_end}
  		
  	\Statex
% 	\lComment{2) scan through $ww\textrm{-}group:$}
	\label{line:init_ww_group_begin}
    \State $lw:$ the last write in $r'_{\Downarrow}$ and on process
    $p(D(r))$
    \State $o_{pre} \gets lw$
	\ForAll{$w \in grp_{ww}$}
 	  \State \Call{PW-Update}{$o_{pre}, w$}; $o_{pre} \gets w$
%  	  \State $o_{pre} \gets w$
	\EndFor
 	\State \Call{PW-Update}{$o_{pre}, r$} 
%  	\Comment{update \textsl{PW} of $r$}	
 	\label{line:init_ww_group_end}	
	\EndProcedure
  \end{algorithmic}
\end{algorithm}
%%%%%%%%%%%%%

%%%%%%%%%%%%%%%%%
\subsubsection{Procedure \textsc{Apply-Rule-C}} 
Procedure \mbox{\textsc{Apply-Rule-C}} is called once the reachability
relation has been dynamically updated. 
Basically it applies Rule~C if necessary and returns false if some cycle is
created (Algorithm~\ref{alg:apply_rule_c}).
In the following, we refer to the three operations involved in Rule
C as ``the $w', w, \textrm{ and } r$ parts of Rule~C'' or simply ``$w', w,
\textrm{ and } r$''. 
We also use the term ``$w'wr$ triple''.
% Also, we assume that $w', w, \textrm{ and } r$ are on the same variable $v$.

\emph{First, to identify the $w'wr$ triple of Rule~C (procedure
\mbox{\textsc{Identify-Rule-C}}):} For some $w'$, it is sufficient to check
whether new paths like from $w'$ to $r$ arise. The notation $\textsl{ReachableRead}$ (Definition \ref{def:reachableread}) serves the purpose. 
For $w'$ (on variable $v$) in check, suppose that its first
reachable read operation $\textsl{RR}[w']$ has been changed from $r_{old}$ to
$r_{new}$.
It means that $w'$ can now reach the read operations in $R[r_{new} \ldots
r_{old})$ which denotes the set of read operations between $r_{new}$ and
$r_{old}$ on process $p_0$ (formally, $R[r_{new} \ldots r_{old}) \triangleq
\{ r \in R \mid r_{new} \preceq_{PO} r \prec_{PO} r_{old} \}$)
(Lines~\ref{line:apply_rule_c_identify_triple_rold}~-~\ref{line:apply_rule_c_identify_triple_rnewold}).
For each read operation $r$ on variable $v$ in $R[r_{new} \ldots r_{old})$, a
triple of $w', w = D(r), r$ is identified.
If there are more than one such $r$, we takes the \emph{first} one (in program
order) and its corresponding triple
(Line~\ref{line:apply_rule_c_identify_triple_w'wr}).
This choice is justified in Lemma \ref{lemma:correctness_applyrulec}. 

%%%%%%%%%%%%% procedure: Apply-Rule-C
% \begin{figure}[!t]
\begin{algorithm}[!t]
  \caption{Procedure \textsc{Apply-Rule-C}.}
  \label{alg:apply_rule_c}
  \begin{algorithmic}[1]
%     \lComment{apply Rule C if necessary; return false if some cycle is created.}
    \lComment{$r_{loop}$: the read operation under scrutiny in outer loop}
    \Procedure{Apply-Rule-C}{$w', r_{loop}$}
	\State $w \gets $ \Call{Identify-Rule-C}{$w'$}
% 	\If{$w = \textsl{NILWRITE}$}
% 	  \State \Return true \Comment{no need to apply Rule C}
% 	\EndIf
	\lIf{$w = \textsl{NIL}$}{\Return true}
    \State add edge $w' \to w$
    \lIf{\Call{Cycle-Detection}{$w',w$}}{\Return false} 
    \State \Call{Update-Reachability}{$w',w,r_{loop}$}
    \State \Return true
    \EndProcedure
    
    \hStatex
    \Setlineno{1}
%     \lComment{identify the $w'wr$ triple of Rule~C}
    \Procedure{Identify-Rule-C}{$w'$}
	\label{line:apply_rule_c_identify_triple_begin}
	\State $r_{old} \gets $ the last value of $\textsl{RR}[w']$
	\label{line:apply_rule_c_identify_triple_rold}
    \State $r_{new} \gets \textsl{RR}[w']$
%     \lComment{new read operations which $w'$ can reach now}
    \State $R[r_{new} \ldots r_{old}) \triangleq \{ r \in R \mid r_{new}
    \preceq_{PO} r \prec_{PO} r_{old} \}$
    \label{line:apply_rule_c_identify_triple_rnewold}
%     \lComment{check each read operation in program order}
    \ForAll{$r_{tmp} \in R[r_{new} \ldots r_{old})$ \emph{in program order}} 
%     \Comment{in $\prec_{PO}$ order}
      \If{$var(r_{tmp}) = var(w')$}
        \State $r \gets r_{tmp}$; $w \gets D(r)$; \Return $w$
        \label{line:apply_rule_c_identify_triple_w'wr} 
%         \Comment {$w'wr$ is identified}
%         \State \Return $w$	
%         \Comment{take the first one}
      \EndIf
    \EndFor
    \State \Return \textsl{NIL} \Comment{no such $w'wr$ triple}
    \label{line:apply_rule_c_identify_triple_end}    
    \EndProcedure
    
    \hStatex
    \Setlineno{1}
%     \lComment{whether a cycle involving $w' \to w$ is created}
    \Procedure{Cycle-Detection}{$w',w$}
%     \lComment{a path $w \leadsto \textsl{PW}[w'][var(w')] \leadsto w'$
%     exists}
    \lIf{$w \preceq \textsl{PW}[w'][var(w')]$}{\Return true}
	\label{line:apply_rule_c_cycle_detection_compare} 
%     \State \Return true
% 	\label{line:cycle_detection_compare_end}
%     \EndIf
    \State \Return false
    \EndProcedure 
    
    \hStatex
    \Setlineno{1}
    \Procedure{Update-Reachability}{$w',w,r_{loop}$}
%     \lComment{update \textsl{ReachableRead} of $w'$ to the earlier read}
    \label{line:apply_rule_c_update_rr_begin}
    \State $\textsl{RR}[w'] \gets \min(\textsl{RR}[w'], \textsl{RR}[w])$
    \label{line:apply_rule_c_update_rr}
    \label{line:apply_rule_c_update_rr_end}
    
%     \lComment{update \textsl{PrecedingWrite} of $w$ and its successors}
    \ForAll{$o \in \{o \mid w \preceq o \preceq r_{loop} \}$} 
    \label{line:apply_rule_c_update_pw_begin}
      \State \Call{PW-Update}{$w',o$} 
      \label{line:apply_rule_c_update_pw_end}
    \EndFor 
    \EndProcedure   
  \end{algorithmic}
\end{algorithm}
% \end{figure}
%%%%%%%%%%%%%

\emph{Second, cycle detection (procedure \mbox{\textsc{Cycle-Detection}}):} 
After identifying a $w'wr$ triple of Rule~C and adding the
edge $w' \to w$, 
procedure \mbox{\textsc{Cycle-Detection}} is called to check whether some cycle
involving $w' \to w$ is created.
To complete a cycle with the new edge $w' \rightarrow w$, an existing path from
$w$ to $w'$ (denoted $w \leadsto w'$) is needed.
The notation \textsl{PrecedingWrite} (Definition \ref{def:precedingwrite})
serves the purpose. 
% Note that the precedence relation between write operations are established via
% either program order or $\prec_{W'WR}$ order. 
Note that $w$ (on variable $v$) concerned here has dictated read operations.
$\textsl{PW}[w'][v]$ maintains the \emph{last} write operation
on variable $v$ which precedes $w'$ and also has dictated read operations. 
Thus cycle detection amounts to figuring out whether or not $w$ precedes (or is)
$\textsl{PW}[w'][v]$ (Line~\ref{line:apply_rule_c_cycle_detection_compare}).
% the precedence relation between $\textsl{PW}[w'][v]$ and $w$.
% (Line~\ref{line:cycle_detection_compare_begin}).

% %%%%%%%%%%%%%%%%% procedure: Cycle-Detection
% \begin{figure}[!t]
%   \begin{algorithmic}[1]
%     \Require $w',w:$ there is an edge $w' \to w$ for $\prec_{W'WR}$ order.
%     \Ensure return true if a cycle is detected; false, otherwise.
%     \Procedure{Cycle-Detection}{$w',w$}
%     
%     \lComment{whether a path $w \leadsto \textsl{PW}[w'][var(w')] \leadsto w'$
%     exists}
%     \If{$w \preceq \textsl{PW}[w'][var(w')]$}
%     \label{line:cycle_detection_compare_begin} 
%     \State \Return true	\Comment{a cycle involving $w',w$ is detected}
%     \EndIf
%     \State \Return false
%     \EndProcedure
%   \end{algorithmic}
%   \caption{Procedure \textsc{Cycle-Detection}}
%   \label{alg:cycle_detection}
% \end{figure}
% %%%%%%%%%%%%%%%%%

\emph{Third, to update the reachability relation (procedure
\mbox{\textsc{Update-Reachability}}):}
If no cycle is created, \mbox{\textsc{Update-Reachability}} is called to update
the reachability relation, namely \textsl{ReachableRead} of $w'$ and \textsl{PrecedingWrite} of $w$ and
its successors.
The \textsl{ReachableRead} of $w'$ is updated to the read operation
$\textsl{RR}[w]$ if $\textsl{RR}[w] \prec_{PO} \textsl{RR}[w']$.
% (Line~\ref{line:apply_rule_c_update_rr}).
Note that \mbox{\textsl{ReachableRead}} of $w'$'s predecessors will be updated
in procedure \textsc{Topo-Schedule}.
The \textsl{PrecedingWrite} of $w$ and its successors (in
$r_{loop_{\Downarrow}}$) are updated to integrate that of $w'$.
% (Lines~\ref{line:apply_rule_c_update_pw_begin} - \ref{line:apply_rule_c_update_pw_end}). 
% In addition, $\textsl{ActiveWrites}$ is also updated (Lines
% \ref{line:apply_rule_c_update_aw_begin} -
% \ref{line:apply_rule_c_update_aw_end})
%%%%%%%%%%%%%%%%%%
\subsubsection{Procedure \textsc{Topo-Schedule}}

Recall that procedure \textsc{Topo-Schedule} mainly involves a serial of
applications of Rule~C and returns false once some cycle is created. . 
The key is that the applications of Rule~C are carried out \emph{locally and
in a well-organized order}.
First, the operations which may act as the $w'$ parts of Rule~C are all
locally in $D(r)\textrm{-}$downset.
Second, they are carried out in a \emph{reverse topological order} of
the $D(r)_{\Downarrow}\textrm{-}$induced subgraph.
The former claim follows from a simple argument: 
\begin{inparaenum}[\itshape a\upshape)]
  \item whether to apply Rule~C is determined by \textsl{ReachableRead} of
  its $w'$ part (procedure \mbox{\textsc{Identify-Rule-C}});
  \emph{and}
  \item \textsl{ReachableRead} of $w'$ is updated only due to its
  successors; \emph{and}
  \item the procedure \mbox{\textsc{Topo-Schedule}} is called immediately after
  some Rule~C edges to $D(r)$ are added (Lines
  \ref{line:centric_start_schedule_begin}~-~\ref{line:centric_start_schedule_end}
  of Algorithm~\ref{alg:read_centric}).
\end{inparaenum} 

%%%%%%%%%%%%% procedure: Topo-Schedule
% \begin{figure}[!t]
\begin{algorithm}[!t]
  \caption{Procedure \textsc{Topo-Schedule}.}
  \label{alg:topo_schedule}
  \begin{algorithmic}[1]
%   \Ensure{schedule operations in $r_{\Downarrow}$; return false once
%   some cycle is created.}
%   \lComment{schedule operations in $r_{\Downarrow}$; return false once
%   some cycle is created.}
  \Procedure{Topo-Schedule}{$r_{\Downarrow}$}
    \lComment{data structures for reverse topological sorting}
    \State $\mathcal{G}_{D(r)_{\Downarrow}} \gets D(r)_{\Downarrow}\textrm{-}$
    induced subgraph
    \State traverse $\mathcal{G}_{D(r)_{\Downarrow}}$ to compute for each
     $o \in D(r)_{\Downarrow}$:
    \State (a) COUNT: number of direct successors
    \State (b) SUCLIST: list of direct successors 
    \State (c) PRELIST: list of direct predecessors
    
    \hStatex
    \lComment{queue to maintain ``sink'' operations}
    \label{line:topo_schedule_queue_begin}
    \State QZERO $\gets$ empty queue
    \State enqueue(QZERO, $D(r)$) \Comment{start from $D(r)$}
	\label{line:topo_schedule_queue_end}

    \lComment{schedule in a reverse topological order of
    $\mathcal{G}_{D(r)_{\Downarrow}}$} 
    \label{line:topo_schedule_topo_process_begin}
    \While{QZERO is not empty}
      \State $w' \gets$ dequeue(QZERO)
      \lComment{apply Rule C if necessary}
      \If{$w' \in W \land w'.DONE = false$}
%         \State $r_{old} \gets \textsl{RR}[w']$	
        \ForAll{$o \in w'.$SUCLIST}
          \State $\textsl{RR}[w'] \gets \min(\textsl{RR}[w'], \textsl{RR}[o])$
        \EndFor
        \State $cycle \gets$ \Call{Apply-Rule-C}{$w', r$}
        \lIf{$cycle$}{\Return false}
        \label{line:topo_schedule_topo_process_end}
        
        \hStatex
        \lComment{Rule C is applied; edge $w' \to w$ is added}
        \label{line:topo_schedule_dynamic_begin}
%         \lComment{$w$ has not been marked DONE yet}
        \If{$w \in D(r)_{\Downarrow} \land (w.DONE = false)$}
          \State insert $w'$ into $w.$PRELIST
          \State insert $w$ into $w'.$SUCLIST
          \State $w'.$COUNT $\gets w'.$COUNT + 1
        \EndIf
        \label{line:topo_schedule_dynamic_end}
      \EndIf
      
      \hStatex
%       \lComment{$w'$ is DONE; erase dependency on it}
      \If{$w'.$COUNT = 0} \label{line:topo_schedule_done_begin}
        \State $w'.$DONE $\gets$ true
        \ForAll{$o \in w'.$PRELIST}
          \State $o.$COUNT $\gets$ $o.$COUNT - 1
          \lIf{$o.$COUNT = 0}{enqueue(QZERO,$o$)}
        \EndFor
      \EndIf
      \label{line:topo_schedule_done_end}     
    \EndWhile
    \State \Return true
  \EndProcedure
  \end{algorithmic}
\end{algorithm}
% \end{figure}
%%%%%%%%%%%%% reschedule algorithm %%%%%%%%%

In the following, we show how to organize the applications of Rule~C (Algorithm
\ref{alg:topo_schedule}).
The basic idea is to integrate the applications of Rule~C with a
(reverse) topological sorting algorithm \cite{Cormen09}.
In such a reverse topological sorting algorithm, a queue is used to
maintain the \emph{sink operations} that have no successors (Lines
\ref{line:topo_schedule_queue_begin} - \ref{line:topo_schedule_queue_end}).
Each time we pick up (and remove) one of the sink operations (denoted $w'$),
update its $\textsl{ReachableRead}$ based on its direct successors, and apply
Rule~C if necessary (Lines~\ref{line:topo_schedule_topo_process_begin} -
\ref{line:topo_schedule_topo_process_end}).
After $w'$ has been processed, it is marked DONE and the dependencies on it are
erased. The new sink operations are put into the queue (Lines
\ref{line:topo_schedule_done_begin} - \ref{line:topo_schedule_done_end}).
However, the applications of Rule~C can introduce new edges
into the subgraph $\mathcal{G}_{D(r)_{\Downarrow}}$. 
Suppose now that an edge from $w'$ to $w$ is added.
In particular, it is subtle when $w \in D(r)_{\Downarrow}$ (meaning that
it is possible for $w$ to act as the $w'$ part of Rule~C) and $w$ has not been
marked DONE yet. 
In this case, it is necessary to process $w$ first \emph{before} marking $w'$
DONE.
This is implemented by imposing dependency of $w'$ on $w$ (Lines
\ref{line:topo_schedule_dynamic_begin} - \ref{line:topo_schedule_dynamic_end}).
The efficiency of procedure \mbox{\textsc{Topo-Schedule}} is justified in Lemma
\ref{lemma:topo_schedule}.
%%%%%%%%%%%%%%%%%%%
\subsection{An Illustrating Example} \label{ss:centric_example}

%%%%%%%%%%%%%%%%%%%%%
\begin{figure}[!t]
  \centering
  \includegraphics[width =
  0.48\textwidth]{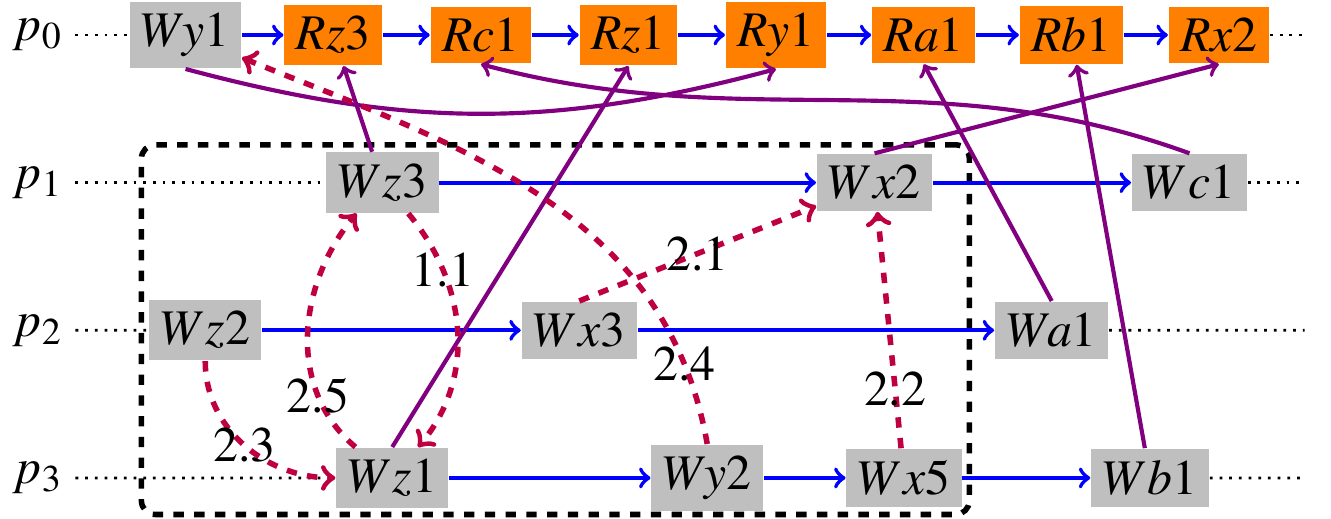}
  \caption{Illustration of the \textsc{Read-Centric} algorithm.}
  \label{fig:ex_schedule_cycle}
  \vspace*{-10pt}
\end{figure}
%%%%%%%%%%%%%%%%%%%%%

Figure~\ref{fig:ex_schedule_cycle} shows a running example of the
\textsc{Read-Centric} algorithm, mainly concerning its sketch and the key
procedure \textsc{Topo-Schedule}.
Assume that $Rx2$ is now under scrutiny (i.e., $r = Rx2$ in Line
\ref{line:centric_foreach} of Algorithm~\ref{alg:read_centric}).
Note that the edge $Wz3 \to Wz1$ (label 1.1) has already been added
due to $Rz1$. 
The \emph{schedule} procedure starts with adding edges $Wx3 \to Wx2$ (label 2.1)
and $Wx5 \to Wx2$ (label 2.2) (Lines~\ref{line:centric_start_schedule_begin} -
\ref{line:centric_start_schedule_end}).
It then calls the procedure \textsc{Topo-Schedule} in the case of
$Wx2 \in Rb1_{\Downarrow}$ (Line~\ref{line:centric_call_schedule}).

In procedure \textsc{Topo-Schedule} (Algorithm~\ref{alg:topo_schedule}), the
operations which may act as the $w'$ parts of Rule~C are in
$Wx2\textrm{-}$downset (in a rectangle dotted box).
Suppose in the course of reverse topological sorting, $Wz2$ is processed
\emph{before} $Wy2$ and $Wz1$.
By Rule~C, an edge $Wz2 \to Wz1$ (label 2.3) is added. 
Since $Wz1$ is not DONE, we have to process $Wz1$ first before marking $Wz2$
DONE (Lines~\ref{line:topo_schedule_dynamic_begin} -
\ref{line:topo_schedule_dynamic_end}).
According to the reverse topological order, $Wy2$ is processed and an edge
$Wy2 \to Wy1$ (label 2.4) is added.
Then it is $Wz1$'s turn. Since there is a path $Wz1 \leadsto Rz3$ via the
edge $Wy2 \to Wy1$, Rule~C is applied and an edge $Wz1
\to Wz3$ (label 2.5) is added.
A cycle involving $Wz1$ and $Wz3$ is thus created.

%%%%%%%%%%%%%%%%%%%%
\subsection{Correctness Proof} \label{ss:centric_correctness}

In this section, we establish the correctness of the
\mbox{\textsc{Read-Centric}} algorithm by showing that it is equivalent to
the \textsc{RW-Closure} algorithm in the sense that their resulting graphs
have the same reachability relation.
Because the edges for both program order and write-to order are static, they
are the same for two algorithms.
The set of $w'wr$ triples identified in the \textsc{Read-Centric} algorithm
is a \emph{subset} of that identified in the \textsc{RW-Closure} algorithm.
The only possible \emph{missing} of $w'wr$ triples is due to procedure
\textsc{Apply-Rule-C}.
% This is justified by the following lemma.

%%%%%%%%%%%%%
\begin{lemma} \label{lemma:correctness_applyrulec}
  In procedure \textsc{Apply-Rule-C}, for $w'$, only the \emph{first} $r$ in
  $R[r_{new} \ldots r_{old})$ is considered for Rule~C (sub-procedure
  \mbox{\textsc{Identify-Rule-C}}).
  This choice does not reduce any reachability relation of the resulting graph
  of the \mbox{\textsc{RW-Closure}} algorithm.
\end{lemma}

  %%%%%%%%%%%%%%
\begin{figure}[!t]
  \centering
	\includegraphics[width = 2.0in]{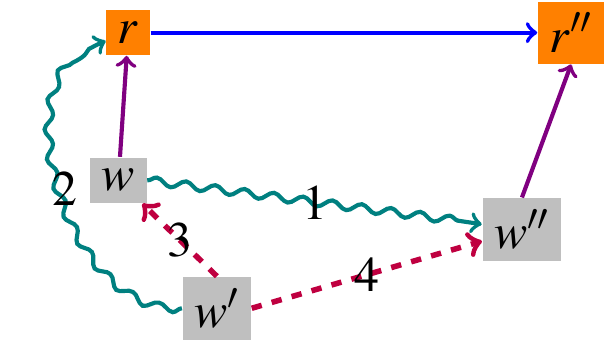}
	\caption[labelInTOC]{Procedure \textsc{Apply-Rule-C} only considers $r$
	  for Rule~C.}
	\label{fig:centric_correctness}
	\vspace*{-10pt}
\end{figure}
  %%%%%%%%%%%%
  
\begin{IEEEproof}
  It is sufficient to show that each missing edge for w'wr order is implied by
  other existing edges.
  This is illustrated in Figure~\ref{fig:centric_correctness} in which all
  operations perform on the same variable and $w = D(r), w'' = D(r'')$.
  For $w'$ there exists a path $w' \leadsto r$ (label 2).
  By Rule~C, both the edge $w' \to w$ (label 3) and the edge $w' \to w''$
  (label 4) should be added.
  However, the latter one is implied by:
  \begin{inparaenum}[\itshape 1\upshape)]
    \item a path $w \leadsto w''$ (label 1) whose existence is guaranteed by $r
    \prec_{PO} r''$; \emph{and}
    \item the edge $w' \to w$ (label 3).
  \end{inparaenum}
\end{IEEEproof}

% According to the above lemma, we have that the \mbox{\textsc{Read-Centric}}
% algorithm is equivalent to the \mbox{\textsc{RW-Closure}} algorithm in the
% sense that their resulting graphs have the same reachability relation. 
Hence, the correctness of the \mbox{\textsc{Read-Centric}} algorithm
follows from that of the \mbox{\textsc{RW-Closure}} algorithm.
% (Theorem~\ref{thm:closure_correctness}).

\begin{theorem} \label{thm:correctness_centric}
  The VPC-MU instance satisfies PRAM consistency if and only if
  the \textsc{Read-Centric} algorithm terminates with a DAG.
\end{theorem}

%%%%%%%%%%%%%%%%%%%%%%%%%%%%%%%%%%%%%%%%%%%%%%%%%%%%%%%%%%%%%%%%%%%%%%%%%%%%%%%%%%%%%
\subsection{Time and Space Complexity} \label{ss:centric_complexity}

The worst-case time complexity of the \mbox{\textsc{Read-Centric}} algorithm
is dominated by the cost of \mbox{\textsc{Topo-Schedule}}.
The efficiency of the latter is justified by the following lemma.

%%%%%%%%%%%%%
\begin{lemma} \label{lemma:topo_schedule}
  Let $r$ be the read operation under scrutiny. 
  For each $w' \in D(r)_{\Downarrow}$, procedure \textsc{Topo-Schedule}
  \emph{applies Rule~C at most once} with it as the $w'$ part. 
\end{lemma}

%%%%%%%%%%%%%%
\begin{figure}[!b]
  \vspace*{-10pt}
  \centering
	\includegraphics[width = 2.0in]{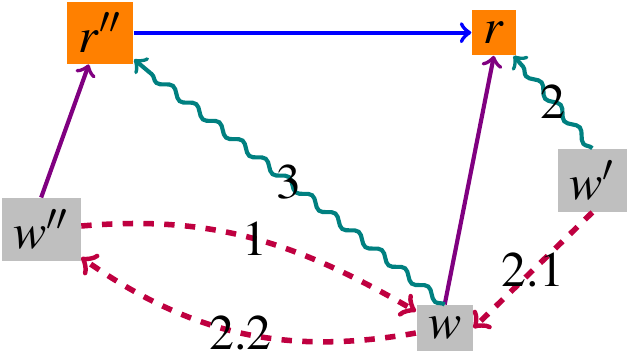}
	\caption[labelInTOC]{Procedure \textsc{Topo-Schedule} applies Rule~C at
	most once for $w'$ .}
	\label{fig:centric_complexity}
\end{figure}
%%%%%%%%%%%%
  
\begin{IEEEproof}
  In procedure \textsc{Topo-Schedule}, the only case in which $w'$ will be
  \emph{checked} for Rule~C more than once is that an edge $w' \to w$ is
  added, $w$ is in $D(r)_{\Downarrow}$, and $w$ has not been marked DONE yet (Lines
  \ref{line:topo_schedule_dynamic_begin} - \ref{line:topo_schedule_dynamic_end}
  in Algorithm~\ref{alg:topo_schedule}).
  In this case, we show that Rule~C is \emph{not applicable} when $w'$ is
  checked again. This is illustrated in Figure~\ref{fig:centric_complexity} in
  which all operations perform on the same variable $v$ and $w = D(r), w'' =
  D(r'')$.
  The \emph{first} application of Rule~C to triple $w', w, \textrm{ and } r$
  have introduced the edge $w' \to w$ (label 2.1). 
  Assume, \emph{by contradiction}, that Rule~C is applicable when $w'$ is
  checked again.
  It requires that via $w$ a new read operation $r''$ on variable $v$ with $r''
  \prec_{PO} r$ be now reachable. 
  Back to the time when $w$ was checked, $r''$ was reachable from $w$ (label 3).
  An edge $w \to w''$ (label 2.2) was added, closing a cycle with
  the edge $w'' \to w$ (label 1) whose existence is guaranteed by $r''
  \prec_{PO} r$.
  The procedure \textsc{Topo-Schedule} would abort then.
\end{IEEEproof}
%%%%%%%%%%%%%

% %%%%%%%%%%%%%
% \begin{table}[!t]
%   \caption{The worst-case time complexity of subroutines in data structures.}
%   \label{tbl:timecomplexity}
%   \centering
%   \begin{tabular}{|c||c|c|}
%     \hline
%     Data structure & Subroutine & \tabincell{c}{Worst-case \\ time complexity}
%     \\ \hline \hline
%     \multirow{3}{*}{GlobalActiveWrites($W_{\mathcal{A}}$)} & replaceWith &
%     $O(1)$ \\ \cline{2-3}
%     & deactivateFrom & $O(n)$ \\ \cline{2-3}
%     & unionWith 	 & $O(n^2)$ \\ \hline
%     \multirow{3}{*}{EarliestRead($r_{\ll}$)} & initEarliestRead & $O(1)$
%     \\ \cline{2-3}
%     & updateEarliestRead & $O(n)$ \\ \cline{2-3}
%     & identifyWRPair & $O(n)$ \\ \hline
%     LatestWrites($W_{\Sigma}$) & updateLatestWrites & $O(\mid V \mid) = O(n)$
%     \\ \hline
%   \end{tabular}
% \end{table}
% %%%%%%%%%%%%%

The following theorem gives the overall worst-case time complexity of the
\textsc{Read-Centric} algorithm.  

\begin{theorem} \label{thm:centric_complexity}
  The worst-case time complexity of the \mbox{\textsc{Read-Centric}} algorithm
  is $O(n^4)$.
\end{theorem}

\begin{IEEEproof}
  Suppose that read operation $r$ is under scrutiny. There are at most $n$
  operations in $r_{\Downarrow}$ and $m = O(n^2)$ edges between them.
  The time complexity of procedure \mbox{\textsc{Topo-Schedule}} comprises
  \begin{inparaenum}[\itshape 1\upshape)]
    \item $O(n + m)$ for reverse topological sorting;
    \item $O(n \cdot c_{apply})$ for at most $n$ applications of Rule~C (Lemma
    \ref{lemma:topo_schedule}), each of which costs:
	\begin{equation*}
      \begin{split}
        c_{apply} = \underbrace{O(n)}_{\textsc{Identify-Rule-C}} &+
        \underbrace{O(1)}_{\textsc{Cycle-Detection}} \\
        &+ \underbrace{O(1 + n + m + n \cdot
        n)}_{\textsc{Update-Reachability}} = O(n^2).
      \end{split}
    \end{equation*}
  \end{inparaenum}
   
  Thus procedure \textsc{Topo-Schedule} costs $O(n^3)$ in the worst case.
  Then the worst-case time complexity of the \mbox{\textsc{Read-Centric}}
  algorithm is $O(n^4)$:
  \[
    \underbrace{O(n)}_{\textrm{iterations}} \cdot\;
    ( \underbrace{O(n^2)}_{\textsc{Init-Reachability}} + 
      \underbrace{O(n^3)}_{\textsc{Topo-Schedule}} ) = O(n^4).
  \]
\end{IEEEproof}

The space complexity of the \textsc{Read-Centric} algorithm is $O(n^2)$:
\[
  \underbrace{O(n)}_{\textsl{ReachableRead}} +
  \underbrace{O(n^2)}_{\textsl{PrecedingWrite}} +
  \underbrace{O(n^2)}_{\textsl{LocalWrites}} = O(n^2).
\]
%%%%%%%%%%%%%%%%%%%%%%%%%%%%%%%%%%%%%%%%%%%%%%%%%%%%%%%%%%%%%%%%%%%%%%%%%%%%%%%%%%%%%
\section{The VPC-SD and VPC-MD Problems are $\sf{NP}$-complete} \label{section:npc}

In this section, we show that the VPC-SD problem (so is VPC-MD) is
$\sf{NP}$-complete by reducing the strongly \mbox{$\sf{NP}$-complete} problem
\textsc{3-Partition} \cite{Garey75, Garey79} to it.

\begin{definition}[\textsc{3-Partition}] \label{def:3-partition} \hspace{3cm}
  \begin{itemize}
    \item \textbf{INSTANCE:} Set $A$ of $3m$ elements, a bound $B \in \Z^{+}$,
    and a size $s(a) \in \Z^{+}$ for each $a \in A$ such that $B/4 < s(a) < B/2$
    and $\sum_{a \in A} s(a) = mB$.
    \item \textbf{QUESTION:} Can $A$ be partitioned into $m$ disjoint sets
    $A_1, A_2, \ldots, A_m$ such that, for $1 \leq i \leq m$, $\sum_{a \in A_i}
    s(a) = B$ (note that each $A_i$ must therefore contain exactly three
    elements from $A$)?
  \end{itemize}
\end{definition}

We choose to reduce from \textsc{3-Partition} because it is \mbox{$\sf{NP}$-complete} even if
the inputs $a \in A$ and $B$ are provided in unary \cite{Garey79}. We use the
\textsc{Unary 3-Partition} problem.

\begin{theorem} \label{thm:npc}
  VPC-SD is $\sf{NP}$-complete.
\end{theorem}

\begin{IEEEproof} \label{proof:npc}
  \emph{VPC-SD is in NP:} Given a schedule of the \mbox{VPC-SD} instance, it is
  straightforward to check whether it is legal by scanning it in polynomial
  time.
  
  \emph{VPC-SD is NP-hard:} To show that VPC-SD is \mbox{NP-hard}, we shall
  give a polynomial reduction from \textsc{Unary 3-Partition} to it. Let
  $A = \{a_1, a_2, \ldots, a_{3m}\}$, $B \in \Z^{+}$ (given in unary), and
  $s(a_1), s(a_2), \ldots, s(a_{3m}) \in \Z^{+}$ (given in unary) constitute an
  arbitrary instance of \textsc{Unary 3-Partition}.
  In the corresponding VPC-SD instance, we assume that integers $a, a', b,
  b', c, c'$ used as variable values are distinct. 
  As in previous sections, $Wxa$ ($Rxa$) denotes the operation of writing
  (reading) value $a$ to (from) variable $x$.
  
\begin{figure}[!t]
  \centering
    \includegraphics[width = 0.48\textwidth]{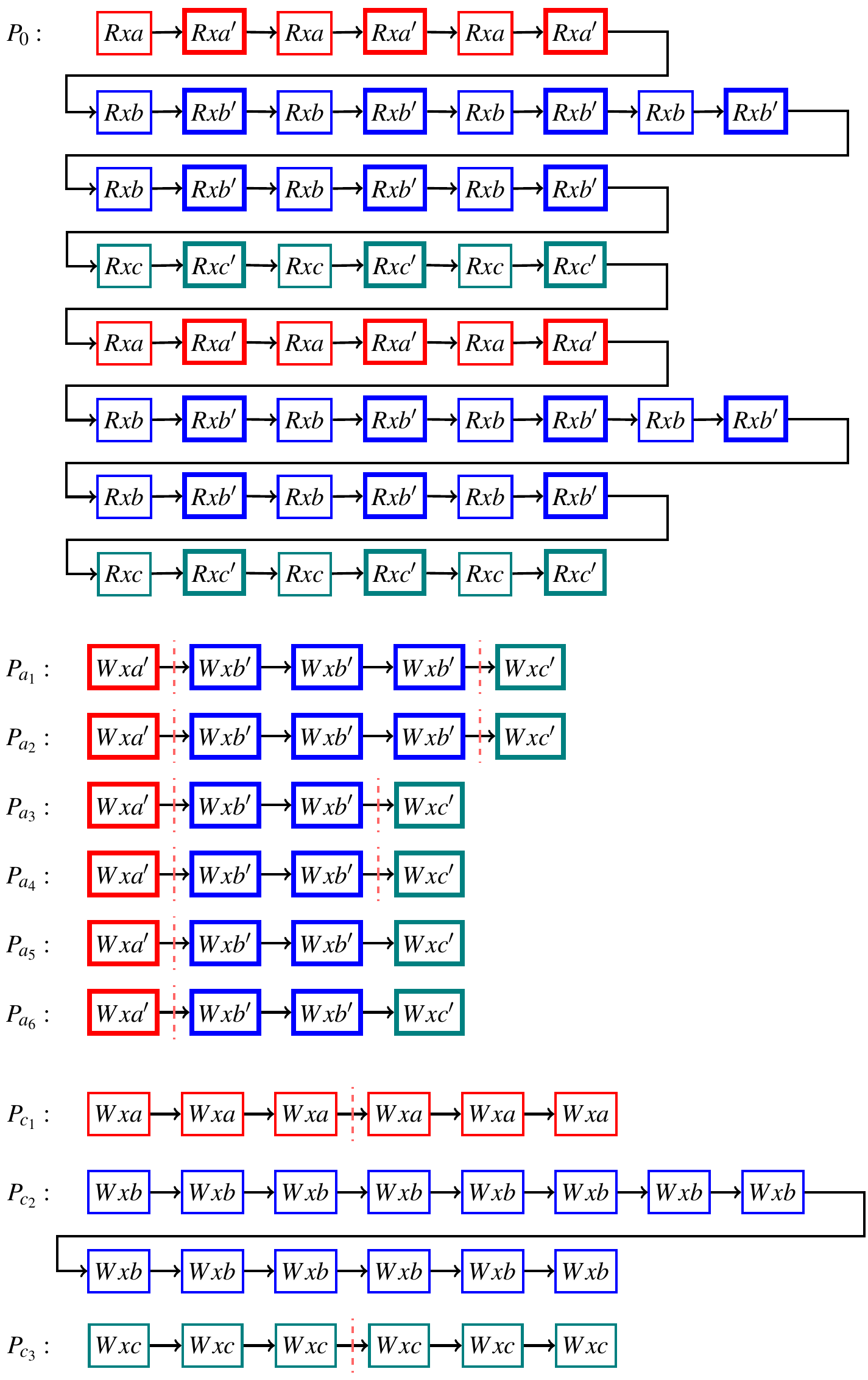}
    \caption{The VPC-SD trace corresponding to an instance of \textsc{Unary
    \mbox{3-Partition}} ($A = \{3,3,2,2,2,2\}, m = 2, B = 7$) obtained with the
    reduction of Theorem \ref{thm:npc}.}
    \label{fig:ICDCS_vpc_sd_npc}
\end{figure}

  The basic idea of the reduction is straightforward: when a schedule encounters
  a read sequence like $Rxa\; Rxa'$, even if the last write of $x$ before the sequence
  is a $Wxa'$, the $Rxa$ forces the schedule to ``use'' another $Wxa'$ to
  satisfy the $Rxa'$.
  
  We represent each $a_i \in A$ with a process $P_{a_i}$ made of $a_i + 2$
  write operations: 
%   the middle $a_i$ write operations $Wxb'$ (blue boxes in figure) are to
%   simulate the element $a_i$ in unary. 
%   To simulate the non-deterministic choice of this element $a_i$, a write
%   operation $Wxa'$ (red boxes) is put in the front of process $P_{a_i}$.
%   Correspondingly, another write operation $Wxc'$ (cyan boxes) is put at the end
%   of process $P_{a_i}$ to complete the choice of this element $a_i$.
  the first operation is a write operation $Wxa'$ (red boxes
  in figure), followed by $a_i$ write operations $Wxb'$ (blue boxes),
  followed by a single write operation $Wxc'$ (cyan boxes).
  
  We then add three auxiliary processes $P_{c_1}, P_{c_2}, P_{c_3}$. 
%   They interact with processes $P_{a_i}$ above to schedule the read
%   sequences like $Rxa\; Rxa'$ ($Rxb\; Rxb'$, and $Rxc\; Rxc'$).
  Specifically, $P_{c_1}$ comprises $3m$ write operations $Wxa$. 
  $P_{c_2}$ comprises $mB = \sum_{a \in A} s(a)$ write operations $Wxb$.
  $P_{c_3}$ comprises $3m$ write operations $Wxc$.
  
  Now we construct the process $P_0$ made only of read operations by
  concatenating $m$ \emph{slot sequences}; each slot sequence is made of:
  \begin{itemize}
    \item a leading \emph{open subsequence} $Rxa$ $Rxa'$ $Rxa$ $Rxa'$ $Rxa$
    $Rxa'$, that forces to pop three operations $Wxa'$ from three distinct $P_{a_i}$
    and open those processes;
    \item followed by a \emph{sum subsequence} $Rxb\; Rxb'$ repeated $B$ times,
    that forces to pop $B$ operations $Rxb'$ from the processes that are currently
    open;
    \item followed by a trailing \emph{close subsequence} $Rxc$ $Rxc'$ $Rxc$
    $Rxc'$ $Rxc$ $Rxc'$, that forces to pop three operations $Wxc'$ from the end of the
    processes that are currently open.
  \end{itemize}
  
  Figure~\ref{fig:ICDCS_vpc_sd_npc} shows an example of the VPC-SD instance
  equivalent to the \textsc{Unary 3-Partition} instance in which $A =
  \{3,3,2,2,2,2\}, m=2, B=7$.

  \emph{The reduction is polynomial:} The size (i.e., total number of
  operations) of the VPC-SD instance is
  \begin{equation*}
    \begin{split}
    \underbrace{(6 + 2B + 6)m}_{P_0} + \underbrace{(6m + Bm)}_{P_{a_i}} &+ 
    \underbrace{3m + Bm + 3m}_{P_{c_1}, P_{c_2}, P_{c_3}} \\ &= 24m + 4Bm.
    \end{split}
  \end{equation*}
  The $a_i$'s and $B$ are given in unary, so it is polynomial in $m$ and
  $B$ and the reduction is polynomial.
  
  We now prove that the \textsc{Unary 3-Partition} instance has a solution if and only if the
  VPC-SD instance has a solution.
  
  ($\Rightarrow$) \emph{If the \textsc{Unary 3-Partition} instance has a solution $A_1,$
  $A_2,$ $\ldots,$ $A_m$,} we construct a legal schedule $\pi$ for the
  \mbox{VPC-SD} instance.
  Let the elements of $A_i$ be $a_{i_1}, a_{i_2}, a_{i_3}$ (in unary). 
  Each $A_i$ corresponds to a subsequence
  $\pi_i$ of $\pi$ in the following way: $P_0$ use the open leading
  subsequence of its $i^{th}$ slot sequence to open each process of
  $P_{a_{i_1}}, P_{a_{i_2}}, \textrm{ and } P_{a_{i_3}}$ by using its $Wxa'$,
  meanwhile ``consuming'' three $Wxa$ from process $P_{c_1}$. The following sum
  sequence completes the $B$ write operations $Wxb'$ from the three currently
  open processes and $B$ write operations $Wxb$ from $P_{c_2}$. Finally, the
  trailing close sequence is scheduled together with $B$ write operations $Wxc'$
  from the three currently open processes and $B$ write operations $Wxc$ from
  $P_{c_3}$. It is straightforward to ensure that the schedule is legal during
  this construction.
  
  ($\Leftarrow$) \emph{If the VPC-SD instance has a legal schedule $\pi$,} we show
  that it is possible to construct a solution to the \textsc{Unary 3-Partition} instance. 
  Note that in $\pi$, read operations and write operations must be scheduled
  \emph{alternately}; otherwise write operations would run out and some read
  operations were left unscheduled. 
  Thus for each slot sequence of $P_0$, $P_0$ has to first use its leading open
  subsequence to open three processes of the $m$ unary $P_{a_i}$. We claim that
  the total number of $Wxb'$ in the three opened processes equals $B$.
  Otherwise, there are two cases:
  \begin{inparaenum}[\itshape 1\upshape)]
    \item the total number of $Wxb'$ is greater than $B$. 
    This means that a process is opened, the corresponding sum subsequence of
    $P_0$ is consumed, and some $Wxb'$ are still there. 
    In order to complete the current trailing close subsequence, we pop them
    (without corresponding $Rxb'$) to reach the final $Wxc'$. 
    However, in one of the next slot sequences there will be not enough $Wxb'$
    to schedule and to reach its close subsequence.
    \item the total number of $Wxb'$ is less than $B$. 
    This means that we are in the middle of a sum subsequence and we need a
    $Wxb'$, but we have already reached the end of all the currently opened
    processes.
    We cannot open another process to recover a $Wxb'$ to complete the sum
    subsequence. Otherwise in one of the next slot sequences there will be not
    enough $Wxa'$ to complete an open subsequence.
  \end{inparaenum}
  
  Thus, VPC-SD is NP-hard and in NP. Therefore VPC-SD is $\sf{NP}$-complete.
\end{IEEEproof}

Note that the largest integer value assigned to the variables in the VPC-SD
instance can be constant (e.g., $a = 1, a'= 2, b = 3, b'= 4, c =
5, c' = 6)$, so it is trivially polynomially bounded by the instance size.
Therefore we can further conclude that VPC-SD is $\sf{NP}$-complete in the strong
sense \cite{Garey79}.
  
Because VPC-MD is a generalization of VPC-SD, we have:
% There is a direct reduction from VPC-SD to VPC-MD: an instance of VPC-SD is also
% a valid instance of VPC-MD, so we have:

\begin{corollary} 
  VPC-MD is $\sf{NP}$-complete.
\end{corollary}
%%%%%%%%%%%%%%%%%%%%%%%%%%%%%%%%%%%%%%%%%%%%%%%%%%%%%%%%%%%%%%%%%%%%%%%%%%%%%%%%%%%%%
\section{Concluding Remarks} \label{section:conclusion}

In this work, we have studied the problem of verifying PRAM consistency
over read/write traces (VPC, for short).
Specifically, we proposed two polynomial algorithms for its VPC-MU variant, 
namely \textsc{RW-Closure} and \mbox{\textsc{Read-Centric}} with the time
complexity $O(n^5)$ and $O(n^4)$, respectively.
We also proved that both its VPC-SD and \mbox{VPC-MD} variants are
$\sf{NP}$-complete.

The verification problems with respect to other weak consistency models, e.g.,
causal consistency \cite{Ahamad95}, are also worth investigation.
Because PRAM is a weakening of causal consistency, our $\sf{NP}$-complete result
also applies to the general problem of verifying causal consistency.
However, it remains open to solve its restricted variant when writes can only
assign unique values for each shared variable.
Moreover, it would be interesting to further study the complexity issues of
evaluating the severity of consistency violations \cite{Golab11,Golab13}.
%%%%%%%%%%%%%%%%%%%%%%%%%%%%%%%%%%%%%%%%%%%%%%%%%%%%%%%%%%%%%%%%%%%%%%%%%%%%%%%%%%%%%
\section*{Acknowledgments}

This work is supported by the National Natural Science
Foundation of China (No. 61272047, 61021062) and the
National 973 Program of China (2009CB320702).
%%%%%%%%%%%%%%%%%%%%%%%%%%%%%%%%%%%%%%%%%%%%%%%%%%%%%%%%%%%%%%%%%%%%%%%%%%%%%%%%%%%%%

\bibliographystyle{IEEEtran}
\bibliography{IEEEabrv,vpc}

%%%%%%%%%%%%%%%%%%%%%%%%%%%%%%%%%%%%%%%%%%%%%%%%%%%%%%%%%%%%%%%%%%%%%%%%%%%%%%%%%%%%%
\end{document}